\documentclass[%
one clumn
amsmath,amssymb,
aps,
pra,
]{revtex4-2}

\usepackage{graphicx}
\usepackage{dcolumn}
\usepackage{epsfig}
\usepackage{graphics,color}
\usepackage{amssymb}
\usepackage{amsmath}
\usepackage{latexsym}
\usepackage{hyperref}
\hypersetup{colorlinks=true, citecolor=blue, urlcolor=blue, linkcolor=red}
\usepackage{amsbsy}
\usepackage{bm}
\usepackage{url}

\begin{document}
\title{Bubble nucleation in the two-flavor quark-meson model}
\author{Junrong Wang}
\author{Ziwan Yu}
\author{Hong Mao}
\email {mao@hznu.edu.cn (corresponding author)}
 \affiliation{1. School of Physics, Hangzhou Normal University, Hangzhou 311121, China}


\begin{abstract}
We investigate the dynamics of a first-order quark-hadron transition via homogeneous thermal nucleation in the two-flavor quark-meson model. The contribution of the fermionic vacuum loop in the effective thermodynamics potential and phase diagram together with the location of critical end point (CEP) have been obtained in the temperature and chemical potential plane. For a weak and strong first-order phase transition, by taking the temperature as a variable, the critical bubble profiles, the evolutions of the surface tension and the saddle-point action in the presence of a nucleation bubble are numerically calculated in detail when fixing the chemical potentials at $\mu=306 \mathrm{MeV}$ and $\mu=309 \mathrm{MeV}$. Our results show that the system could be trapped in the metastable state for a long time as long as the temperature is between the metastable region characterized by the up and low spinodal lines. Moreover, the surface tension at criticality will rise up to about $4 \mathrm{MeV/fm^2}$ when the chemical potential is very high. Such a small value of the surface tension would favor a mixed phase in the cores of compact stars and may have an important implication in astrophysics.
\end{abstract}


\maketitle

\section{Introduction}
It is widely believed that the hadronic matter characterized by confinement and chiral-symmetry at low net baryon densities undergoes a phase transition into a deconfined and chirally symmetric quark-gluon-plasma (QGP) through a smooth crossover as the temperature is increased. At large densities, some studies predict a first-order phase transition line separating the hadronic matter from the QGP and possible more exotic phases. At end of this line, there should exist a so-called critical endpoint (CEP) where the transition should be a continuous second-order one. To investigate and identify phase diagram is one of the most challenging problems in high energy physics and astrophysics \cite{Yagi:2005yb,Fukushima:2010bq,Braun-Munzinger:2015hba}, and the study is experimentally supported by the heavy-ion collision experiments, such as the Relativistic Heavy Ion Collider (RHIC) at Brookhaven National Laboratory and the Large Hadron Collider (LHC) at CERN. These conducted experiments provide us with the opportunity to inspect and reveal fundamental properties of the strong interaction. Moreover, to explore a wider range of the QCD phase diagram up to several times the normal nuclear-matter density, the new Facility for Antiproton and Ion Research at Darmstadt, the Nuclotron-based Ion Collider Facility at the Joint Institute for Nuclear Research in Dubna, and the Japan Proton Accelerator Research Complex at Japan Atomic Energy Research Institute and Japan’s National Laboratory for High Energy Physics have been scheduled and planed, and the CEP can be explored in phase II of Beam Energy Scan program at RHIC and in upcoming experiments \cite{Luo:2017faz,Pandav:2022xxx}. 

From a theoretical point of view, Quantum Chromodynamics (QCD), the gauge theory describing strong interactions in elementary particle physics, is applicable for determining the properties of strongly interacting matter at finite temperature and density. However, due to the fermion sign problem, an \textit{ab initio} approach, Lattice Field Theory, is severely hampered by the failure of importance sampling if a chemical potential is involved \cite{Aarts:2023vsf}. In order to describe the low-energy nonperturbative phenomena in the framework of QCD theory, an alternative approach is effective models with possessing two salient features of QCD, i.e. chiral symmetry and confinement. To mention a few, these effective models, which are successfully utilized for many decades, are the Nambu-Jona-Lasinio (NJL) model \cite{Nambu:1961tp,Nambu:1961fr}, the linear sigma model (LSM) \cite{Gell-Mann:1960mvl} and their modernized extensions, the Polyakov Nambu-Jona-Lasinio model (PNJL) \cite{Costa:2010zw,Fukushima:2017csk} and the Polyakov Quark Meson Model (PQM) \cite{Schaefer:2007pw,Mao:2009aq,Schaefer:2009ui}.

Recently, after the discovery of gravitational waves by LIGO Collaboration \cite{LIGOScientific:2016aoc}, the subject of cosmological first-order phase transition has gathered increasingly interest due to the stochastic gravitational wave background which would be produced \cite{Caprini:2019egz,Hindmarsh:2020hop,Croon:2023zay}. The stochastic gravitational wave background could be detected by current and near future detectors. Once this signal is to be observed, it would provide us with the earliest known probe of the universe. Moreover, aside from the early universe phenomena in primordial first-order phase transition, the observation of gravitational waves also shed light on the field of astrophysics, future gravitational wave observations related to a first-order quark-hadron phase transition would not only enable the probing of the equation of state for matter under extreme circumstances, but also give a constrain on the quark-hadron surface tension. In combination with other observations, astrophysics has now entered the multimessenger era \cite{Paschalidis:2017qmb,Sieniawska:2018zzj,Han:2018mtj,Xia:2019pnq,Baiotti:2019sew,Blacker:2020nlq,Cao:2018tzm}. Therefore, to understand the dynamics of the first-order phase transition is crucial and important. With upcoming gravitational waves experiments, there is a need to explore the anticipated phenomenology tightly connected with its underlining fundamental mechanism.    

It is well known that the dynamics of first-order phase transition in the early universe and heavy-ion collisions at ultrarelativistic energies can be applied through the homogeneous nucleation theory \cite{Csernai:1992tj,Kapusta:2006pm}. The modern theory pioneered by Langer in the late 1960s in the context of classical statistical mechanics \cite{Langer:1967ax,Langer:1969bc} has been extended  to relativistic quantum field theory by Callan and Coleman for zero temperature \cite{Coleman:1977py,Callan:1977pt,Coleman:1988} and by Affect \cite{Affleck:1980ac} and Linde \cite{Linde:1980tt,Linde:1981zj} for finite temperature. The remarkable goal of a nucleation theory is to calculate the nucleation rate of a bubble or droplet of a stable (true) vacuum inside a metastable (false) vacuum near the critical temperature. Suppose that a system nearby its critical temperature, due to the thermal and quantum fluctuations of any thermodynamic systems, bubbles of the stable vacuum created by fluctuations may grow or shrink inside the homogeneous false vacuum depending on its energy budget with regard to the false vacuum. If a droplet is too small, the free energy gain from the phase transition of the bulk is less than the energy cost in the creation of an interface between two vacua. The total free energy is positive, the droplet will shrink and evaporate. On the other hand, if the droplet is large, a bulk free energy gain is relatively large and the surface energy cost is negligible, the droplet will tend to grow and eventually occupy the whole system, completing the phase conversion.    

For a strong first-order phase transition, which is usually characterized by an effective potential with a zero-temperature potential barrier, the dynamics of the quark-hadron phase conversion based on the Friedberg-Lee (FL) model \cite{Friedberg:1976eg} has been studied numerically \cite{Zhou:2020bzk}, and the found has been also compared to its analytic results obtained with the thin-wall approximation \cite{Shu-Ying:2022wqb}. Since the FL model is lack of chiral symmetry, the model only predicts a first-order phase transition in whole QCD phase diagram, this, of course, conflicts with other studies based on Lattice simulations or chiral models, so that the model can merely serve as the prototypical toy model for current interests. To fixed this problem, it is necessary to introduce a chiral symmetry in the FL model in order to provide a proper description of hadron-quark phase transition beyond the first-order transition. Then, the quark meson model treated as an upgrade to the FL model seems fulfill the requirements in both the studies of the static nucleon properties and the QCD phase transition \cite{Mao:2013qu,Jin:2015goa}. In the framework of the quark meson model, homogeneous bubble nucleation has been initially investigated both in numerical and analytic methods in Ref \cite{Scavenius:2000bb}, but the study was constrained on the temperature below the critical temperature and an unphysical coupling constant was chosen in order to enhance the strength of the first-order phase transition. Furthermore, with the thin-wall approximation, bubble nucleations at low temperature but high density and the strong magnetic field have been previously investigated in Refs.\cite{Palhares:2010be,Kroff:2014qxa}. In a first-order phase transition, a key ingredient is the surface tension, by using the analytical method or the thin-wall approximation, surface tensions and phase diagram have been obtained in the quark meson model with the Polyakov-loop in Refs \cite{Mintz:2012mz,Stiele:2016cfs}.

In this work, we will carry on a systematic and complete study on the dynamics of the first-order quark-hadron phase transition through the exact numerical method based on the following three important points. The first, and most important, point is that the thin-wall approximation is not applicable when the temperature goes far from the critical temperature, since the radius of the bounce is comparable with the thickness of the bubble wall and the friction force term can be discarded. The situation will become worse for a weak first-order phase transition, due to the fact that the first-order phase conversion will turn from a homogeneous bubble nucleation to a spinodal decomposition when the temperature approaches to the spinodal line, the thin-wall approximation should break down accordingly. So that the exact numerical method is crucial and necessary. In the following discussion, we will compare our results with the recently found in the quark meson model within the thin-wall approximation \cite{Palhares:2010be,Kroff:2014qxa,Mintz:2012mz,Stiele:2016cfs,Wan-Ping:2019ifh}. The second point is that the contribution of the sea quark in the pressure is usually ignored in previous studies. But it plays a significant role in the study of the hadron quark phase transition because it will soften the first-order phase transition and reduce the surface tension of hadron quark interface dramatically \cite{Skokov:2010sf,Gupta:2011ez,Herbst:2013ail}. Moreover, in the chiral limit, the inclusion of the fermionic vacuum term can influence the order of the chiral phase transition in the QM and PQM model and give out a second-order phase transition which is in agreement with the universality argument based on the $O(4)$ symmetry \cite{Skokov:2010sf} . In this work, we will include the fermion vacuum fluctuation in the effective potential in order to present a more realistic phase diagram and QCD thermodynamics. Finally, the surface tension of the interface between two vacua plays a center role in the procedure of a first-order phase transition for the sake that it is the amount energy per unit area costed in the creation of the interface between quark phase and hadron phase. So that lots of current studies and efforts have been drawn to this problem. However, the saddle-point action evaluated on the bounce solution is also important in determining the nucleation rate of the true vacuum inside the homogeneous false vacuum. Moreover, a first-order phase transition is not a fast conversion, it does not happen exactly at the critical temperature. To give out the specific moment when the phase conversion is to be completed, or more specifically, the phase boundary of the quark phase or the hadron phase, we need to compute this action precisely, especially when the temperature is very close to the spinodal temperature. Therefore, in this work, by adopting the realistic coupling constant the saddle-point action will be calculated when the temperature is below the critical coexistence line as well as the temperature is above the critical value.

The structure of the paper is as follows. In the next section we briefly describe the quark meson model. After that we will discuss the effective potential at finite temperatures and densities and present a phase diagram of the QCD phase transition. In Sec. IV, we give detailed description of homogeneous nucleation and the methods used for both numerical and analytic computations of the critical bubble profiles. Our results and discussions are presented in Sec. V, while in the last section we give our conclusions.

\section{The Model}

In terms of chiral fields, the Lagrangians of two massless noninteracting quarks $u$ and $d$ are invariant under the global $SU(2)_L\times SU(2)_R$ chiral phase transformations
\begin{equation}
\psi _{L,R}\rightarrow \psi _{L,R}'=U_{L,R}\psi _{L,R},
\end{equation}
where $\psi_{L,R}=\binom{u}{d}_{L,R}$ and $U_{L,R}=\exp (-i\vec{\theta}_{L,R}\cdot \frac{\vec{\tau}}{2})$ . However this chiral symmetry does not appear in the low energy particle spectrum and the strong interaction theory exhibits the phenomenon of spontaneous symmetry breaking. Consequently, three Goldstone bosons appear and the constituent quarks become massive at low energy. In describing the symmetries of the Lagrangian, it is useful to introduce three pion mesons $\vec{\pi }$ and a $\sigma $ meson in terms of a matrix field as
\begin{equation}
\Phi =\sigma \frac{\tau ^0}2+i\vec{\pi }\cdot \frac{%
\vec{\tau }}2,
\end{equation}
where $\tau ^0$ is the unity matrix and $\vec{\tau }$ are the three Pauli matrices. Under the $SU(2)_L\times SU(2)_R$ chiral symmetry
transformations, $\Phi$ transforms as
\begin{equation}
\Phi \rightarrow \Phi'= U_{L}\Phi U_{R}^{+}.
\end{equation}

Then the renormalizable effective Lagrangian of the two-flavors quark meson model is defined as \cite{Gell-Mann:1960mvl,Scavenius:2000qd}
\begin{equation}\label{Lagall}
{\mathcal{L}}={\mathcal{L}}_\Phi +{\mathcal{L}}_q,
\end{equation}
where
\begin{equation}\label{Lagmeson}
{\mathcal{L}}_\Phi =\textrm{Tr}[(\partial _\mu \Phi )^{+}(\partial
^\mu \Phi )]-\lambda [\textrm{Tr}(\Phi ^{+}\Phi )-\frac{{\vartheta}^{2}
}{2}]^2-H \textrm{Tr}[\Phi],
\end{equation}
and
\begin{equation}\label{Lagquark}
{\mathcal{L}}_q=\overline{\psi }_Li\partial\!\!\!/ \psi _L+\overline{\psi }  
_Ri\partial\!\!\!/\psi _R-2g\overline{\psi }_L\Phi \psi
_R+h.c..
\end{equation}
Here we have introduced a flavor-blind Yukawa coupling $g$ of the left-handed and right-handed quark fields to interact with the $\Phi$ field. 

The parameters of the Lagrangian $\mathcal{L}$ are chosen by the requirement that the chiral symmetry $SU(2)_L\times SU(2)_R$ is spontaneously broken down to $SU(2)_{L+R}$ in the vacuum while the $\sigma$ field takes on a non-vanishing vacuum expectation value $\langle\sigma\rangle ={\it f}_{\pi}=93 \mathrm{MeV}$. It results in a massive $\sigma $ meson and three massless Goldstone bosons $\vec{\pi }$ mesons in chiral limit, as well as giving an effective mass $m_q=gf_\pi $ to the constituent quarks. Furthermore, the chiral symmetry is explicitly broken by adding the last term in Eq.(\ref{Lagmeson}) due to the finite current quark masses. With this additional term, the vector isospin $SU(2)$ symmetry remains exact but the axial $SU(2)$ transformation is no longer invariant. Accordingly, the constant $H$ is to be fixed by the partially conserved axial vector current relation which gives $H=f_{\pi}m_{\pi}^{2}$, where the pion mass is taken as $m_{\pi}=138$ MeV. Moreover, the dimensionless coupling constant $g$ in the model is determined by the constituent quark mass in vacuum, which is about $1/3$ of the nucleon mass and gives $g\simeq 3.3$. The another dimensionless coupling constant $\lambda$ is usually fixed by the sigma mass $m_\sigma^2=m^2_\pi+2\lambda f^2_\pi$, here we set to $500$ MeV in accord with the most recent compilation of the Particle Data Group \cite{ParticleDataGroup:2022pth}. Finally, the quantity $\vartheta$ is actually not a free parameter and can be formally expressed as $\vartheta^{2}=f^{2}_{\pi}-m^{2}_{\pi}/\lambda$.

\section{Effective potential and phase structure}

A convenient framework of studying phase transitions and the restoration of the chiral symmetry at extreme high energy is the thermal field theory \cite{Croon:2023zay,Mustafa:2022got}. Within this framework, the effective potential is one of the important and powerful theoretical tool, and the standard approach for dealing with the thermodynamics of various observables of interest relies on the grand canonical ensemble. To make things lucid, we start with a spatially uniform system in thermodynamical equilibrium at temperature $T$ and quark chemical potential $\mu$, from here and henceforward, we will use the chemical potential to represent the quark chemical potential. In general, the grand partition function is commonly given by
\begin{eqnarray}
\mathcal{Z}= \int\prod_a \mathcal{D} \sigma \mathcal{D} \pi_a \int
\mathcal{D}\psi \mathcal{D} \bar{\psi} \mathrm{exp} \left[ \int_X
(\mathcal{L}+\mu \bar{\psi} \gamma^0 \psi )
\right],
\end{eqnarray}
where $\int_X\equiv \int^{\beta}_0 d\tau \int d^3x$, the inverse temperature $\beta=1/T$, and $\mu=\mu_B /3 $ for the homogeneous background field.

In the mean-field approximation, the meson fields in the Lagrangian are replaced by their expectation values, whereas the quark and antiquark fields are still retained as quantum fields. This implies that the one-loop correction to the effective potential from the quark fields are taken into account, but treats the mesonic degrees of freedom at tree level. Following this scheme, the integration over the fermions yields a determinant which can be calculated by standard procedures \cite{Kapusta:2006pm,Laine:2016hma}, it will generate an effective potential for the mesons. Finally, the effective potential of the model can be obtained exactly in a closed form as
\begin{eqnarray}\label{potential}
\Omega(T,\mu)=\frac{-T \mathrm{ln}
\mathcal{Z}}{V}=U(\sigma ,\vec{\pi} )+\Omega_{\bar{\psi}
\psi},
\end{eqnarray}
where the classical potential for the $\sigma$ and $\vec{\pi}$ is rewritten as
\begin{equation}
U(\sigma ,\vec{\pi})=\frac{\lambda}{4} \left(\sigma ^{2}+\vec{\pi} ^{2}
-{\vartheta}^{2}\right)^{2}-H\sigma,
\label{mpot}
\end{equation} 
and the contribution of quarks and antiquarks are given by
\begin{eqnarray}\label{qeffpot}
\Omega_{\bar{\psi} \psi} &=& \Omega_{\bar{\psi} \psi}^\mathrm{v}+\Omega_{\bar{\psi} \psi}^{th} \nonumber \\ 
 &=& -\nu \int \frac{d^3\vec{p}}{(2
\pi)^3} E \nonumber \\&& 
 -\nu T \int \frac{d^3\vec{p}}{(2
\pi)^3} \left\{ \mathrm{ln} \left[ 1+e^{-(E-\mu)/T}\right] +\mathrm{ln} \left[
1+e^{-(E+\mu)/T}\right]\right\}.
\end{eqnarray}
Here, $\nu=2N_f N_c=12$ and $E=\sqrt{\vec{p}^2+m_q^2}$ is the valence quark and antiquark energy for $u$ and $d$ quarks, and the minus sign is the consequence of Fermi-Dirac statistics. The constituent quark (antiquark) mass is set to $m_q=g\sigma$.

The first term of Eq.(\ref{qeffpot}) denotes the fermions vacuum one-loop contribution which is ultraviolet divergent and can only be evaluated in the presence of a regulator. The divergence in Eq.(\ref{qeffpot}) then can be appropriately renormalized by using the dimensional regularization scheme \cite{Quiros:1999jp,Skokov:2010sf,Gupta:2011ez}. After taking into account the vacuum fluctuations and therefore renormalization issues, the renormalized fermion vacuum one-loop contribution reads
\begin{eqnarray}\label{omegareg}
\Omega_{\bar{\psi} \psi} ^{\mathrm{v}}=\Omega_{\bar{\psi} \psi} ^{\mathrm{reg}}=-\frac{N_c N_f}{8\pi^2} m_q^4 \mathrm{ln}(\frac{m_q}{\Lambda}),
\end{eqnarray}
where $\Lambda$ denotes the arbitrary renormalization scale. It is worth pointing out that dimensional regularization really introduces an arbitrary renormalization scale parameter, nevertheless, at least in the one-loop approximation, the thermodynamic potential and all physical observable are independent of the choice of $\Lambda$, and the scale dependence can be neatly cancelled out after the rearrangement of parameters in the model \cite{Jin:2015goa,Skokov:2010sf,Gupta:2011ez,Li:2018rfu}.

Equipped with the above effective potential, we can explore the phase diagram of the model at finite temperature and density by minimizing the thermodynamical potential in equation (\ref{potential}) with respect to the order parameter $\sigma$. Then an equation of motion is given by
\begin{eqnarray}
\frac{\partial\Omega(T,\mu)}{\partial \sigma}=0.
\end{eqnarray}
The solution of the equation of motion determines the behavior of the chiral order parameters $\sigma$ as a function of $T$ and $\mu$, as well as the phase diagram of the model. As we know, the thermodynamical state of equilibrium is given by the values of the order parameter at the global minimum of the effective potential, once the order parameter for each given $T$ and $\mu$ is obtained, any thermodynamical quantity of equilibrium, such as the pressure, the entropy density, the energy density, the speed of sound, et al., can be described and calculated. 

\begin{figure}
\includegraphics[scale=0.36]{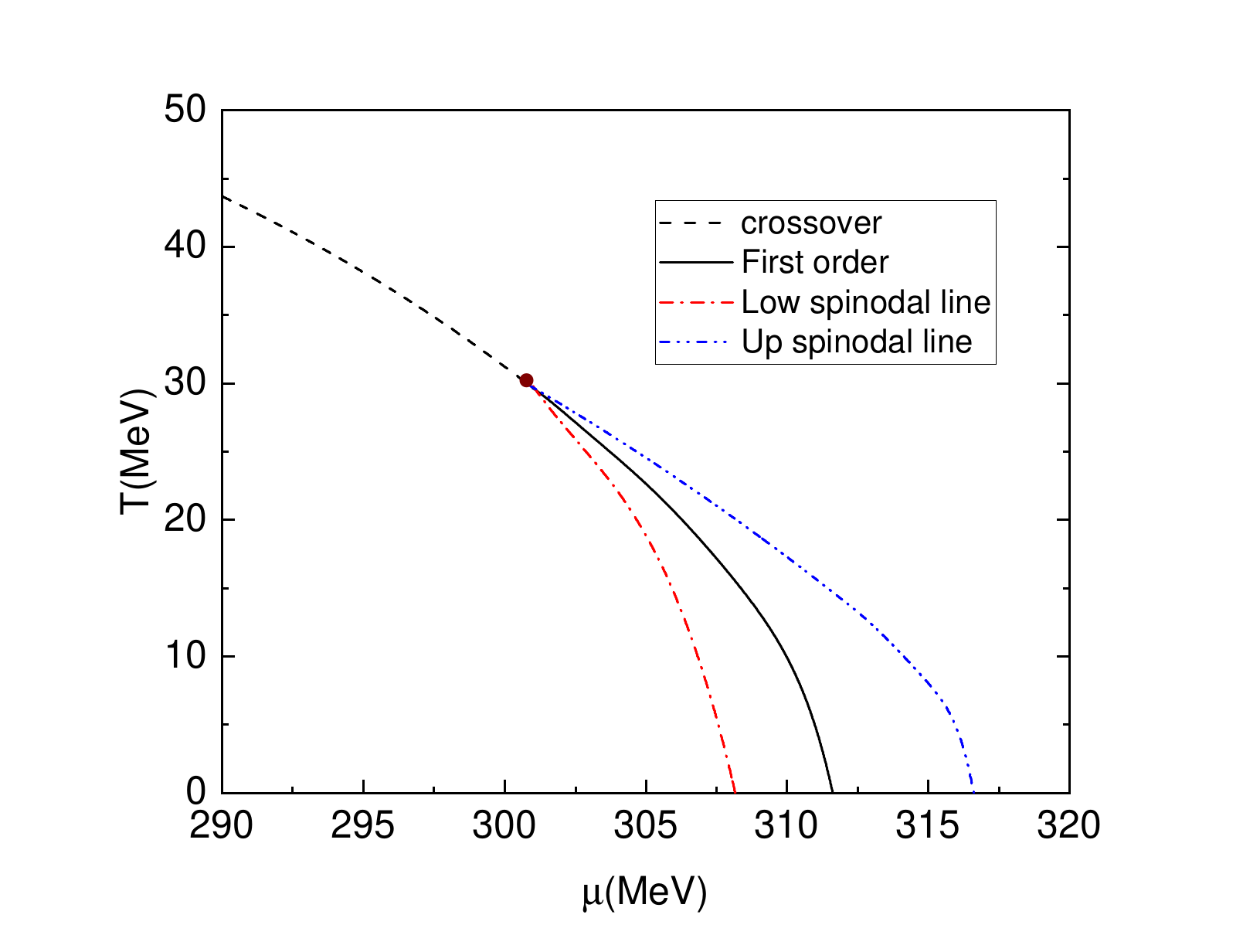}
\caption{\label{Fig01} (Color online) The phase diagram in the $T-\mu$ plane for the two-flavor quark meson model. The dashed lines are the critical line for conventional chiral phase transition in the region of crossover. The solid lines indicates the first-order phase transitions, and the solid circle indicates the critical end points for chiral phase transitions of $u$ and $d$ quarks. The dashed-dotted line and the dashed-doted-dotted line are the lower and upper spinodal lines.}
\end{figure} 

In figure \ref{Fig01}, we have presented the phase diagram in calculation with the fermion vacuum fluctuation for the two-flavor quark meson model. The temperature behavior of the chiral condensate $\sigma$ shows that the system experiences a smooth crossover transition at low chemical potential, while there is a first-order phase transition for larger chemical potential due to the fact that the chiral order parameter makes a jump across the gap of the condensate near the critical temperature $T_c$. Normally, the temperature derivative of the chiral condensate $\sigma$ for quarks has a peak at some specific temperature, which is established as the critical temperature for the chiral phase transition. Because the temperature derivative of the chiral condensate has simply one peak, we can not tell when and where the crossover phase transition would convert to a first-order one at the critical end point (CEP) with a second order phase transition \cite{Wan-Ping:2019ifh,Scavenius:2000qd}. In order to locate the CEP in the phase diagram, the quark number susceptibility $\chi_q=\partial^2 \Omega(T,\mu)/ \partial^2 \mu$ is to be introduced, and it is believed to be divergent at the CEP \cite{Luo:2017faz,Pandav:2022xxx}.

Aside from calculation of the quark number susceptibility $\chi_q$, in the present work, we prefer to use the shapes of the effective potential at various temperatures and chemical potentials to decide the position of the CEP. In the case of the first-order phase transition, along the critical line with the temperature $T\simeq T_c$, the thermodynamical potential $\Omega(T,\mu)$ has two minima of equal depth separated by a potential barrier. With the reduction of the chemical potential, the height of the barrier decreases and finally disappears at the CEP, in which the phase transition is of second order. In our calculation the corresponding CEP is located at $(T_E, \mu_E)\simeq (30, 301)$ MeV in Fig.\ref{Fig01}. It is worth to pointing out that the location of CEP from theory calculations is scattered over region of $\mu_B=200-1100$ MeV and $T=40-180$ MeV \cite{Luo:2017faz,Pandav:2022xxx}. QCD-based model calculations like NJL model \cite{Chen:2015dra,Fan:2016ovc}, QM model \cite{Friman:2011pf,Friman:2014cua}, PNJ \cite{Costa:2010zw} and PQM \cite{Schaefer:2007pw,Mao:2009aq,Schaefer:2009ui} give out a relative larger critical chemical potential around $\mu=\mu_B/3=300$ MeV. However, the functional renormalization group (FRG) approach and Dyson-Schwinger equations predict a rather narrow region for the critical chemical potential among $mu=200-220$ MeV \cite{Gao:2020qsj,Fu:2021oaw,Fu:2023lcm,Fu:2022gou}. Whereas the accuracy of predictions for CEP from the first principle lattice-QCD calculations worsens towards very large chemical potential, various model calculations wildly vary in their predictions. Therefore, experimental search of critical point is crucial and important to establish its position in phase diagram. 

\begin{figure}[thbp]
\epsfxsize=9.0 cm \epsfysize=6.5cm
\epsfbox{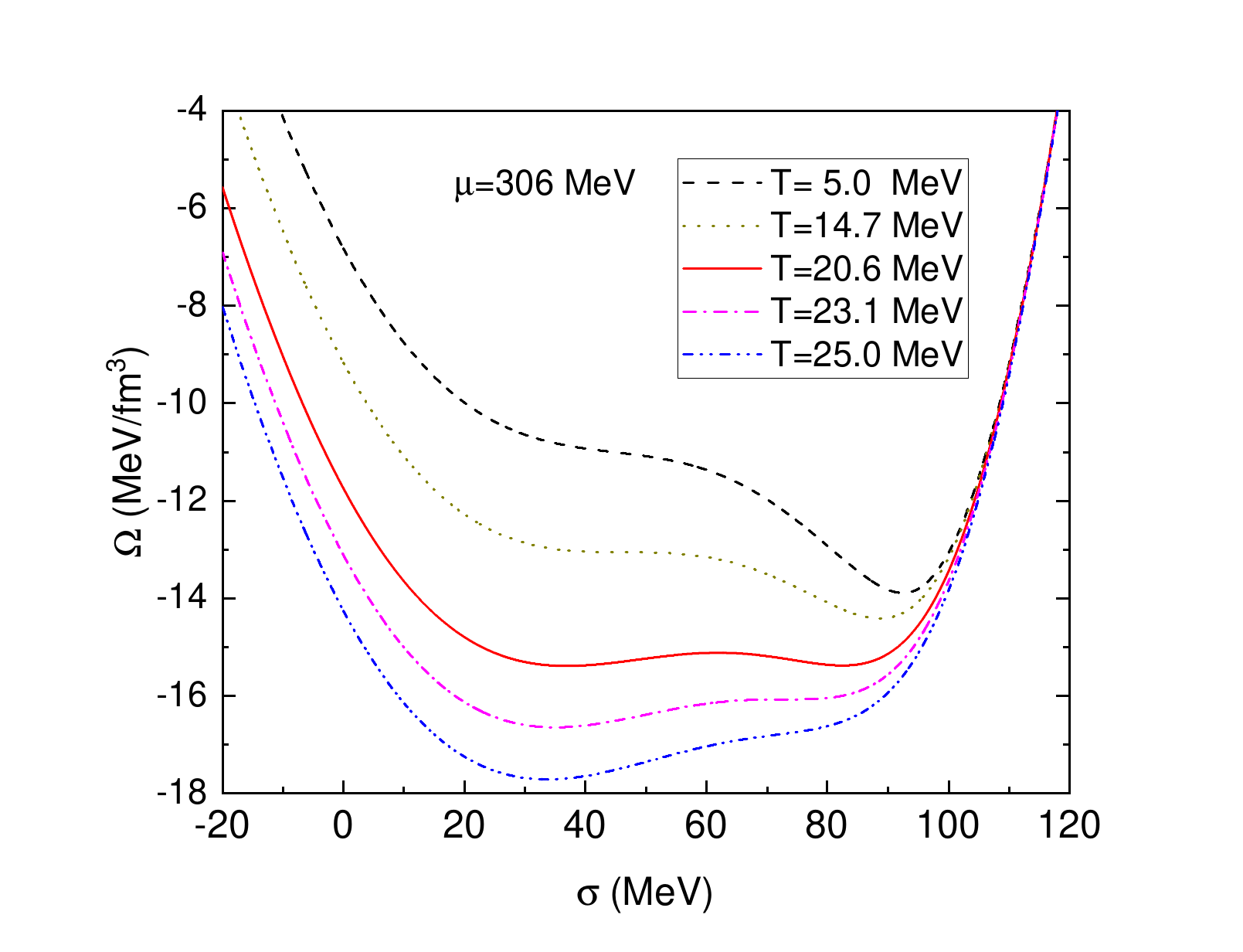}\hspace*{0.01cm} \epsfxsize=9.0 cm
\epsfysize=6.5cm \epsfbox{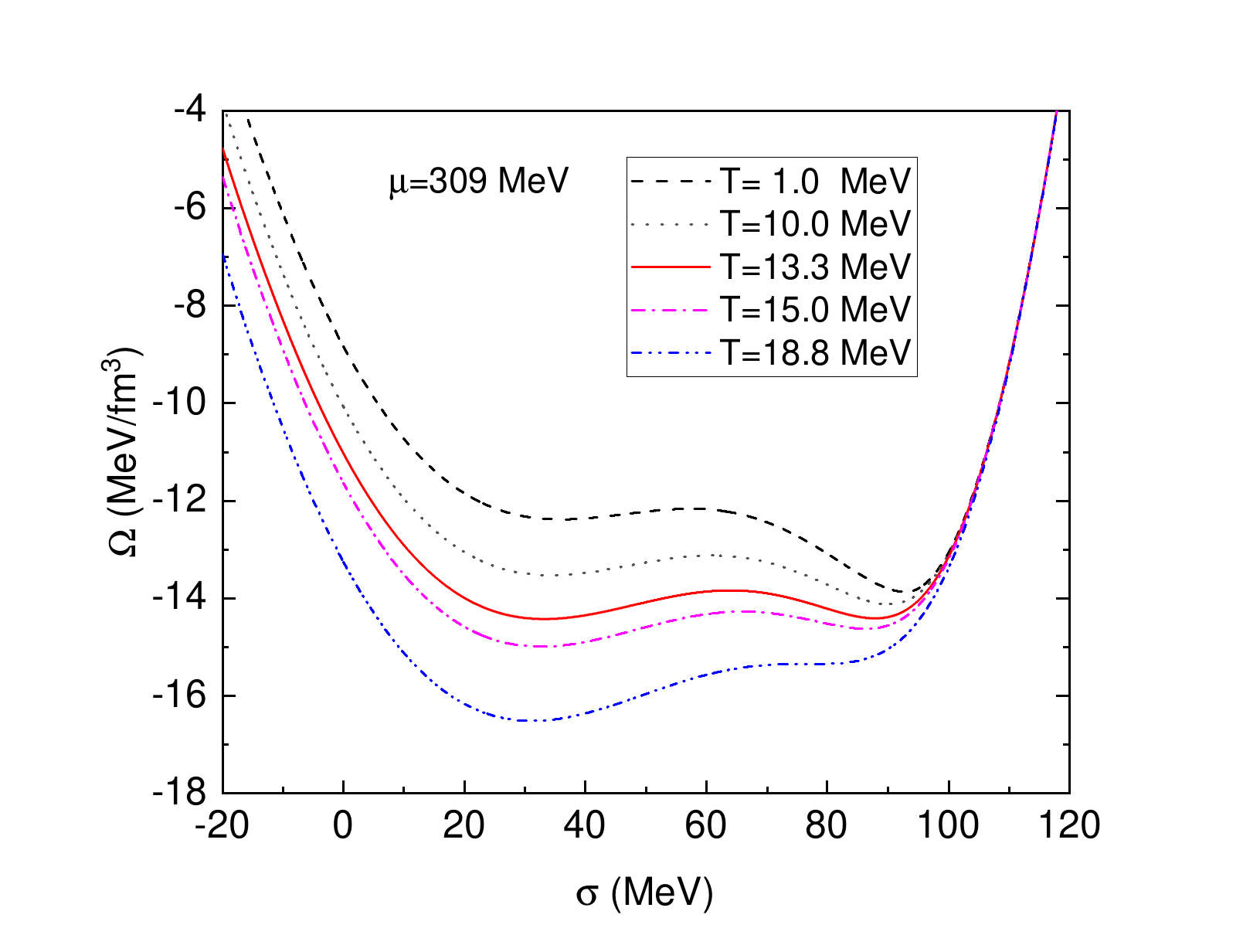}
 \caption{(Color online) (a) The grand canonical potentials $\Omega$ as a function of the chiral order parameter $\sigma$ for $\mu=306$ MeV at various temperatures. (b)The grand canonical potentials $\Omega$ as a function of the chiral order parameter $\sigma$ for $\mu=309$ MeV at various temperatures.}
\label{Fig02}
\end{figure}

As shown in Fig.\ref{Fig02}, in the region of the first-order phase transition, a typical effective potential commonly displays a local minimum at a low sigma $\sigma_l$ which is separated by a potential barrier from another local minimum at a relative larger sigma $\sigma_h$. When a critical temperature $T_c$ is reached, these two minima are degenerate. For $T<T_c$, the minimum of the effective potential at $\sigma=\sigma_h$ is the absolute or global minimum, which is regarded as the stable (true) vacuum, whereas the minimum at $\sigma=\sigma_l$ is treated as the metastable (false) vacuum. In this case the chiral symmetry is broken so that the constituent quarks become massive. On the contrary, when the temperature $T$ goes across above the critical value $T_c$, these two vacua will flip over, the global minimum is now at $\sigma=\sigma_l$, and the local minimum is at $\sigma=\sigma_h$. Since the chiral symmetry is approximately restored and the quarks become almost massless, the system for $T>T_c$ is  then  considered as the quark phase. The previous case for $T<T_c$ is taken as the hadron phase, therefore the critical lines divide the whole phase diagram into two categories: hadron phase and the quark phase.
  
Normally, apart from the critical temperature $T_c$, there have two other temperatures of interests in a first-order phase transitions. These two temperatures $T_{c1}$ and $T_{c2}$ are named as the lower and upper spinodal critical points, respectively. A typical example is shown in Fig.\ref{Fig02}, where the evolutions of the potential for several temperatures when fixed a chemical potential at $\mu=306$ MeV and $\mu=309$ MeV are exhibited. For the left panel in Fig.\ref{Fig02} as $\mu=306$ MeV, when the temperature is around $T_c\simeq20.6$ MeV, the shape of the potential exhibits two degenerate minima. However, with the increase of the temperature, the second minimum of the potential at $\sigma=\sigma_h$ disappears at a higher temperature $T_{c2}\simeq 23.1$ MeV. In the meanwhile, when the temperature is to fall below the critical temperature $T_c$, the first minimum of the potential at $\sigma=\sigma_l$ tends to wipe out around $T_{c1}\simeq 14.7$ MeV. Between these two specific temperatures, metastable states or false vacuum exist, and the system can exhibit supercooling or superheating. 

For $\mu=309 \mathrm{MeV}$, one can also observe the characteristic pattern of a first order phase transition: two minima corresponding to phases of restored and broken chiral symmetry are separated by a potential barrier and they will become degenerate when the temperature is at $T_c \simeq 13.3$ MeV. Chiral symmetry is approximately restored for $T>T_c$, where the minimum at false vacuum $\sigma=\sigma_l$ becomes the absolute minimum as shown in the right panel in Fig.\ref{Fig02}. Similarly to the previous case for $\mu=306$ MeV, when the temperature $T$ goes across up the critical line and rises further, the potential barrier between two minima will start to decrease gradually and shrink to zero at the moment when the second minimum of the potential at $\sigma=\sigma_h$ vanishes at a spinodal temperature $T_{c2}\simeq 18.8$ MeV. On the other hand, when $T<T_c$, the shapes of the effective potential at various low temperatures display quite different behaviors in comparison with those of the previous case as $\mu=306$ MeV. The barrier between two minima of the effective potential is to maintain even when the temperature $T$ is very close to zero. This means that the first minimum of the effective potential at $\sigma=\sigma_l$ always could exist in hadron phase. Therefore, the phase transition could be identified as a strongly first-order phase transition which is usually induced by an effective potential with a nonvanishing zero-temperature potentia barrier.    

To give out a complete description of a first-order phase transition, two particular lines of the spinodal points which constrain the regions of spinodal instability for the first-order phase transition at high density are illustrated in Fig.\ref{Fig01}. Similarly to the critical line in Fig.\ref{Fig01}, both the lower and upper spinodal lines grow up with the reduction of the chemical potential $\mu$, but the gaps of these two spinodal lines become small and small, in the end, two spinodal lines and the critical line will terminate by the same point at the CEP. Moreover, since the lower spinodal line will end up at some point $\mu_c\simeq 308$ MeV on the vertical axis of the chemical potential, the area of the first-order phase transition can be technically split into a weakly first-order phase transition and a strongly first-order phase transition according to the above discussion. 

Therefore, for a weakly first-order chiral phase transition as the chemical potential is $\mu<\mu_c$ as $T<T_c$, the thermodynamic potential exhibits a local minimum aside from the global minimum, when the temperature decreases from $T_c$ to a specific value $T_{c1}$. The local minimum gradually disappears at a point of the inflection known as spinodal instability. Whereas, for $\mu>\mu_c$, the chiral phase transition is to be considered as a strongly first-order one due to the fact that the local minimum remains for the temperature is at $T<T_c$ and there is no spinodal temperature anymore. The critical chemical potential for a transition from a weak first-order phase transition to that of a strong one is then identified as the critical chemical potential at $\mu_c\simeq 308$ MeV in hadron phase \cite{Zhou:2020bzk,Bessa:2008nw}.   

\section{Homogeneous thermal nucleation}

The mechanism of the nucleation theory can be used to study the probability that a bubble or droplet of the stable vacuum in a system initially trapped in the metastable vacuum near the critical temperature $T_c$. For a pure system, the formation of bubbles originates from intrinsic thermodynamic fluctuations, this kind of nucleation mechanism is commonly called as homogeneous nucleation. On the contrary, when impurities cause the formation of bubbles or droplets, such a mechanism of the nucleation theory is known as heterogeneous nucleation. In the everyday world, the external agents would play the role of nucleating centers, such as dust or ions in the atmosphere, leading to a much more efficient increase of the nucleation rate. Nevertheless, for the physical interests related to our study, homogeneous nucleation theory is appropriate and we will use this basic theoretical apparatus to describe the decay of the a metastable vacuum of a system interacting with a heat bath at temperature $T$. 

Based on the framework of the homogeneous thermal nucleation, we make an assumption in the limit that thermal fluctuations dominate quantum fluctuations and the quantumly-induced tunneling is simply ignored. Then, the nucleation rate per unit time per unit volume is given by the form of 
\begin{equation}\label{nuclrate}
\Gamma=\mathcal{P}\exp\left[ -\frac{S_3}{T}  \right],
\end{equation} 
where $T$ is the temperature of the system in equilibrium with the thermal bath, $S_3$ is the three-dimensional action associated with the $O(3)$-symmetric critical bubble or droplet and $\mathcal{P}$ is the exponential prefactor. For the mechanism of the bubble nucleation, to the leading order, the nucleation rate is controlled by the exponent of the three-dimensional action evaluated on the critical bubble. The sub-leading corrections to the leading-order bubble action are included in the prefactor $\mathcal{P}$, which can be technically expressed as
\begin{equation}\label{nuclrate2}
\mathcal{P}= \frac{\omega_-}{2\pi}\left( \frac{S_3}{2\pi T} \right)^{3/2}\left[ \frac{\mathrm{Det}(-\nabla^2+\Omega''_{FV})}{\mathrm{Det}'(-\nabla^2+\Omega''_{B})} \right]^{1/2}.
\end{equation} 
Here $\omega_-$ is the eigenvalue of the negative mode, the terms $\Omega''_{FV}$ and $\Omega''_{B}$ are abbreviations for $\Omega''$ evaluated in the false vacuum and the critical bubble, the prime in the determinant signifies that the zero eigenvalues associated with the translation symmetry of the bubble are omitted. $\Omega''$ is the second derivative of the effective potential $\Omega(T,\mu)$ with respect to the order parameter $\sigma$ which is actually represented a field in describing the extremum of the three-dimensional Euclidean action, more specifically, the critical bubble or the bounce. Usually, the calculation and evaluation of this prefactor is a nontrivial matter, a rough estimate of their ratios can be obtained by dimensional analysis and it can be approximately expressed as $T^4$ or $T^4_c$ for simplicity \cite{Csernai:1992tj,Scavenius:2000bb}. 

The result represented in the above equation (\ref{nuclrate}) is a semi-classical contribution based on a saddle point approximation around the bounce solution. By taking the scalar field $\sigma$ as the order parameter, at finite-temperature field theory, an Euclidean action we are interested in is 
\begin{equation}\label{Euaction}
S_E(\sigma)=\int_0^{\beta}d \tau \int d^3r \left[ \frac{1}{2} \left(\frac{\partial \sigma}{\partial \tau} \right)^2+\frac{1}{2} \left(\nabla \sigma \right)^2+\Omega(\sigma;T,\mu) \right],
\end{equation}
in which the subscript $E$ denotes Euclidean and the integral is over Euclidean space. For the sake of convenience, in the following discussions, we will keep the $\sigma$ field in the effective potential $\Omega$ in Eq.(\ref{potential}) explicitly. As argued by Linde \cite{Linde:1981zj}, for sufficiently high temperature as length scales large compared to $\beta$, the relevant number of dimensions is $d=3$, and the Euclidean action becomes
\begin{equation}\label{Euaction2}
S_E(\sigma)\equiv\frac{S_3}{T},
\end{equation} 
where $S_3$ is the three-dimensional saddle-point action associated with the formation of a critical-sized bubble or droplet, in what follows it is to be called as the saddle-point action for abbreviation. Therefore, the bounce is an $O(3)$ symmetric solution to the classical equation of motion that extremizes the Euclidean action $S_3$. In particular, for a scalar field $\sigma$, the bounce satisfies a nonlinear ordinary differential equation,
\begin{equation}\label{eom}
\frac{d^2\sigma(r)}{dr^2}+\frac{2}{r}\frac{d\sigma(r)}{dr}=\frac{\partial\Omega(\sigma;T,\mu)}{\partial \sigma},
\end{equation}
with boundary conditions $\lim\limits_{r \to \infty }\sigma(r)=\sigma_{FV}$ and $\frac{d\sigma(r)}{d r}|_{r=0}=0$. The first boundary condition is because that the bubbles are embedded in the homogeneous false vacuum, outside the bubble, the $\sigma$ field should arrive at its false vacuum at $\sigma \simeq \sigma_{FV}$. While the second one is set by the requirement of finite energy at the origin. The solution for this equation of motion with the above proper boundary conditions is a saddle point solution or a bounce $\sigma_b$. It is an $O(3)$ non-trivial field configuration that starts in the false vacuum and reaches the other side of the potential barrier with zero velocity. In this work we will use AnyBubble package \cite{Masoumi:2016wot} to determine the bounce.

Once the solution $\sigma_b$ is obtained, the $S_3$ exponent in Eq.(\ref{nuclrate}) can be evaluated on the bounce solution $\sigma_b$ as
\begin{equation}\label{Eu3action}
S_3=\int d^3r \left[ \frac{1}{2} \left(\nabla \sigma \right)^2+\Omega(\sigma;T,\mu) \right],
\end{equation}
and the surface tension of the nucleation bubble interface between the false vacuum and the true vacuum is defined accordingly as \cite{Linde:1981zj,Weinberg:2012pjx}
\begin{equation}\label{surfacet}
\Sigma=\int dr \left[ \frac{1}{2} \left(\frac{d\sigma}{dr}  \right)^2+\Omega(\sigma;T,\mu) \right].
\end{equation}
It is worth noting that in the practical calculations, if the false vacuum has a non-zero potential energy, an additional term $-\Omega(\sigma_{FV};T,\mu)$ should be included in the $S_3$ action and the surface tension $\Sigma$.

For a generic effective potential, the equation of motion of the bounce with boundary conditions usually cannot be obtained analytically, we should rely on numerical methods to do the computation. However, when the system is very close to the critical coexistence line, the bubble radius $R$ is much larger than the wall thickness ($\triangle R \sim m_{\sigma}^{-1}$), hence the damping force $2 \sigma'/r $ in the field equation has become negligible, the thin-wall approximation is applicable and the equation of motion (\ref{eom}) reduces to the field equation for a typical one-dimensional soliton
\begin{equation}\label{tweom}
\frac{d^2 \sigma (r)}{dr^2}=\frac{d \Omega}{d \sigma}.
\end{equation}
This static field equation implies that
\begin{equation}\label{tweom2}
\frac{d \sigma(r)}{dr}=\pm \sqrt{2 \Omega}.
\end{equation}
Integrating Eq.(\ref{tweom2}) yields
\begin{eqnarray}\label{twradius}
  r &=& \int_{\sigma}^{\sigma_v} \frac{d \sigma}{\sqrt{2 \Omega}}.
\end{eqnarray}
Therefore an approximate solution for the bounce can be obtained for arbitrary potential with two or more degenerate minima. Moreover, within the thin-wall approximation, the surface tension of the bubble can be calculated as
\begin{equation}\label{twsurface}
\Sigma_{\mathrm{tw}}=\int_{0}^{\infty} dr \left[ \frac{1}{2} \left(\frac{d\sigma}{dr} \right)^2+\Omega \right]=\int_{0}^{\sigma_v} d\sigma \sqrt{2 \Omega},
\end{equation}
and the saddle-point action $S_3$ is given by
\begin{equation}\label{tw3action}
S_3=\frac{16 \pi}{3} \frac{\Sigma_{\mathrm{tw}}^3}{\varepsilon}.
\end{equation}
The quantity $\varepsilon=\Omega(\sigma_f;T,\mu)-\Omega(\sigma_t;T,\mu)$ is the difference between the values of the effective potential at the false vacuum and at the true vacuum.

\section{Results and discussion}

In this section, we will numerically solve the equation of motion in Eq.(\ref{eom}) with some boundary conditions, $\sigma\rightarrow \sigma_{FV}$ as $r\rightarrow \infty$ and $\sigma'(0)=0$. Since a first-order phase transition necessitates a discontinuity in the scalar field $\sigma$, the transition does not take place exactly at the critical temperature $T_c$ along to the critical coexistence line in phase diagram in figure \ref{Fig01}. Consequently, the false vacuum $\sigma_{FV}$ should be addressed properly before we obtain the exact numerical solutions of the equation of motion. 

In order to get an intuitional description of the false vacuum decay at finite temperature and density, the dynamics of a first-order phase transition at metastability is to be classified into two categories: top-down and bottom-up. More specifically, by ``top-down”, we mean that we will study the false vacuum decay at finite temperature and density for the temperature $T< T_c$ and the initial false vacuum of the system is well prepared as quark phase at $\sigma=\sigma_l$ for $T\simeq T_c$ when the system is cooling down from very high temperature. On the other side, by “bottom-up”, we mean that the original false vacuum for the temperature $T\simeq T_c$ is defined as the hadron phase at $\sigma=\sigma_h$, and we will study the decay of the hadron phase for the temperature $T> T_c$ in the event that the system is heating up from the low energy state.    

\subsection{Top-down}

\begin{figure}[thbp]
\epsfxsize=9.0 cm \epsfysize=6.5cm
\epsfbox{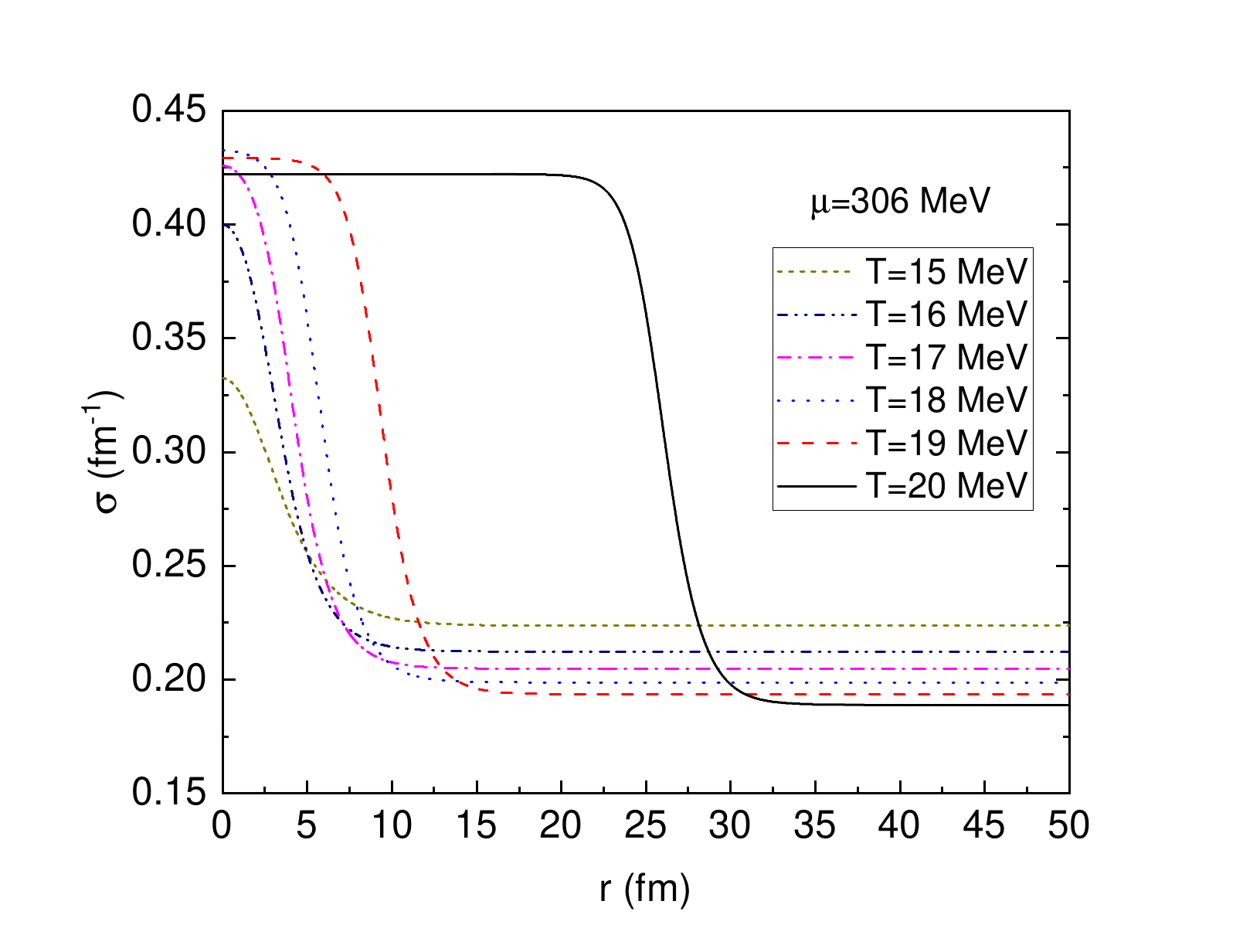}\hspace*{0.01cm} \epsfxsize=9.0 cm
\epsfysize=6.5cm \epsfbox{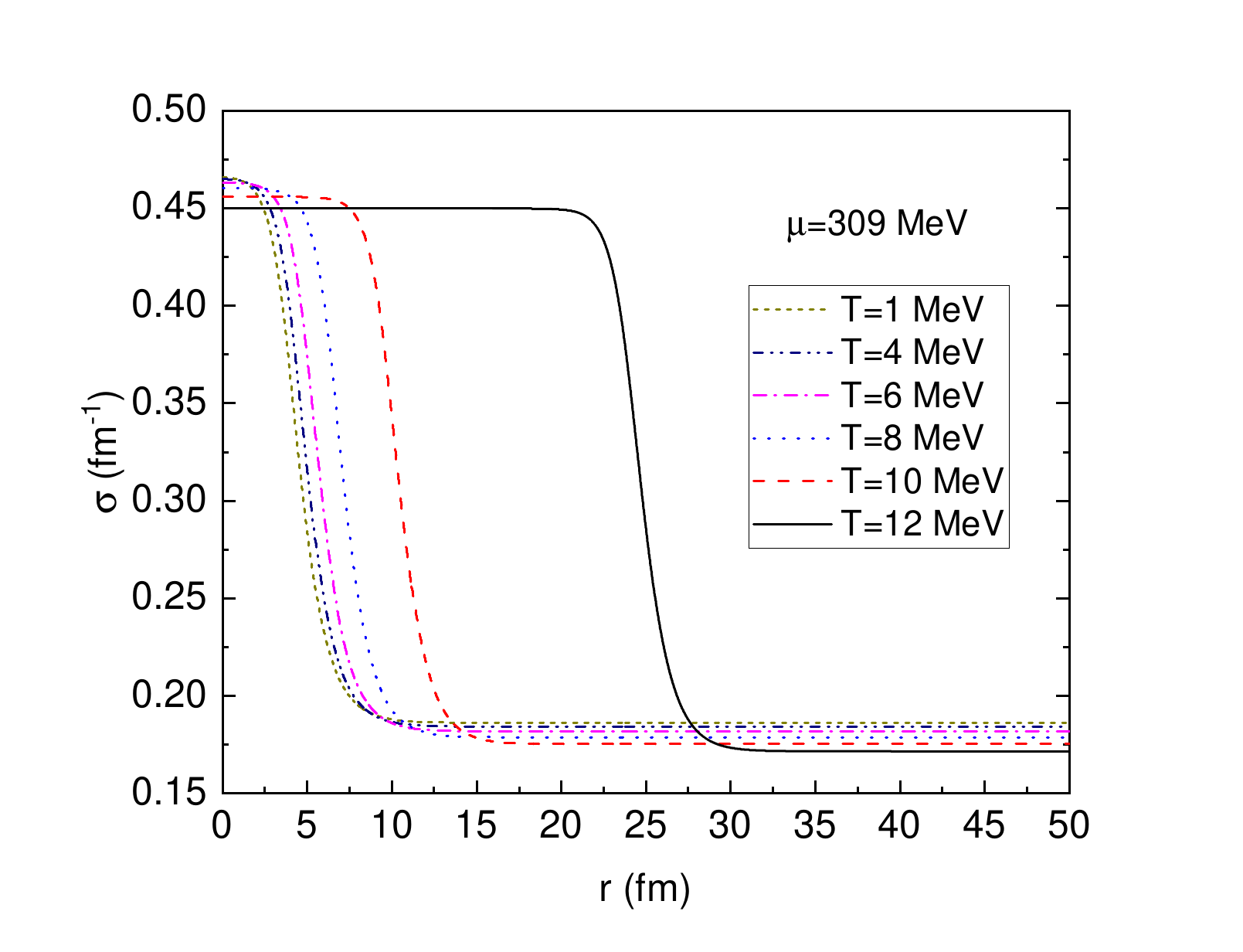}
 \caption{(Color online) (a) Critical bubble profiles for different temperatures when fixing the chemical potential at $\mu=306$ MeV for $T<T_c$. From left to right, the curves correspond to $T=15$, $16$, $17$, $18$, $19$ and $20$ MeV. (b) Critical bubble profiles for different temperatures when fixing the chemical potential at $\mu=309$ MeV for $T<T_c$. From left to right, the curves correspond to $T=1$, $4$, $6$, $8$, $10$ and $12$ MeV.}
\label{Fig03}
\end{figure}

As the system is cooling down from very high temperature. When $T\simeq T_c$, the initial false vacuum can be causally set as the quark phase since the first-order transition does not happen exactly at $T_c$ but upon lowing the temperature. The appearance of a bubble of the hadron phase (the stable state) inside the quark phase (the metastable state) is a natural consequence of the thermal fluctuations of the thermodynamical system sufficiently close to the coexistence line in phase diagram. To study the dynamics of a first-order phase transition in this ``top-down" scenario as $T<T_c$, we numerically solve the equation of motion in Eq.(\ref{eom}) with the specific boundary conditions as $\sigma\rightarrow \sigma_l$ as $r\rightarrow \infty$ and $d \sigma(0)/dr=0$. Here, the false vacuum $\sigma_l$ is temperature-dependent as a matter of fact that the local minimum of the effective potential varies with increasing of the temperature when fixing the chemical potential.

For $\mu=306$ MeV, the exact numerical solutions by taking the temperatures as $T=15$, $16$, $17$, $18$, $19$, $20$ MeV are plotted in the left panel of Fig.\ref{Fig03}.  It is shown that with the temperature decreasing from $T_c\simeq20.6$ MeV, all curves approach to their false vacuum $\sigma_l$ when the radius $r$ is large, whereas $\sigma(r)$ at the center of the bubble deviates from its stable vacuum value at $\sigma=\sigma_v$ significantly. When the temperature is sufficiently close to the critical temperature, the $\sigma$ field at the center of the bubble only slightly deviates from its stable vacuum value at $\sigma=\sigma_h$, however, as soon as $T < T_c$ MeV, the $\sigma(0)$ field is visibly different from its stable vacuum value. Such a deviation can be qualitatively explained by an ``overshoot-undershoot" argument due to Coleman \cite{Coleman:1977py}. According to this argument, the equation of motion (\ref{eom}) is reinterpreted as the equation for a particle moving in an ``upside-down" potential energy $\Omega$, the second term in the field equation is taken as a damping force. When the system is very close to the critical coexistence line, the bubble radius $R$ is very larger, so that the damping force can be neglected, the field $\sigma$ can start to roll down at the top of the effective potential $\Omega$ around $\sigma\simeq\sigma_h$ to rest at its false vacuum $\sigma_l$. However, with the temperature descends, the radius of the bubble decreases accordingly, the damping force in the field equation becomes important, consequently, the field $\sigma(0)$ will deviate from its true vacuum value more and more dramatically. In other words, the thin-wall approximation mentioned above is expected to be invalid, and any further extension of the thin-wall approximation to the temperatures deviation from $T_c$ should be checked very carefully.
 
A similar discussion can be applied to the second case when the chemical potential is fixed at $\mu=309$ MeV. The critical bubble profiles at different temperatures are illustrated in the right panel of Fig.\ref{Fig03}, where the temperatures are taken as $T=1$, $4$, $6$, $8$, $10$, $12$ MeV from left to right. The evolution of $\sigma(r)$ for different temperatures tells that the typical radius of the critical bubble should increase as well as the increase of the temperature, and $\sigma(r)$ approaches to its false vacuum value at $\sigma=\sigma_l$ as $r\rightarrow \infty$. Also, from this figure, as long as the temperature is lower than the critical temperature $T_c$, $\sigma(0)$ will depart from its stable vacuum value at $\sigma=\sigma_v$ significantly, this nontrivial behavior of the $\sigma(r)$ in the center of the bubble can also be interpreted as a limit to the applicability of the thin-wall approximation. 

\begin{figure}[thbp]
\epsfxsize=9.0 cm \epsfysize=6.5cm
\epsfbox{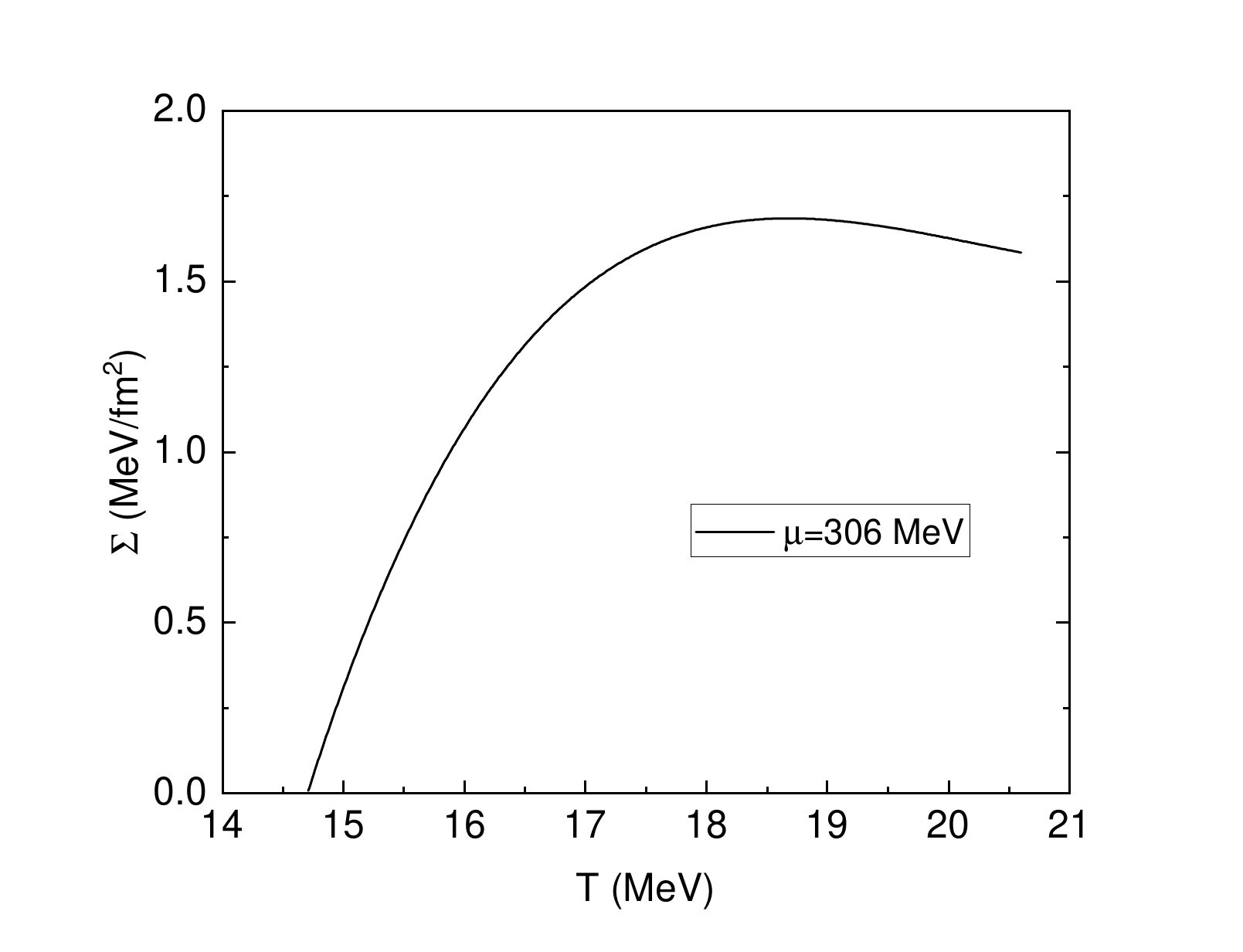}\hspace*{0.01cm} \epsfxsize=9.0 cm
\epsfysize=6.5cm \epsfbox{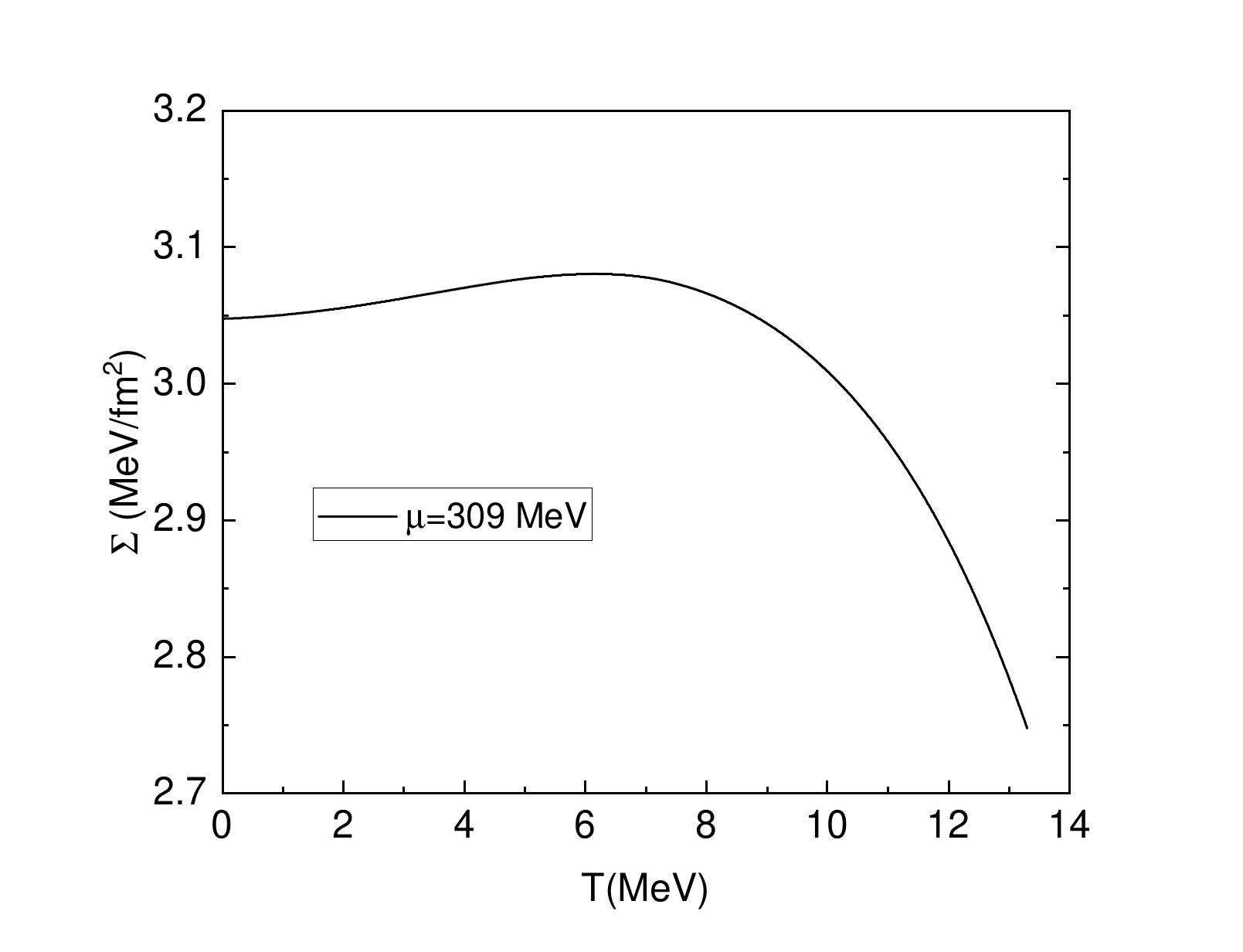}
 \caption{(Color online) (a) Surface tension as a function of temperature $T$ for $T\leq T_c$ at $\mu=306$ MeV. (b) Surface tension as a function of temperature $T$ for $T\leq T_c$ at $\mu=309$ MeV.}
\label{Fig04}
\end{figure}

Once the bubble profiles have been solved, from the definition in Eq.(\ref{surfacet}), the surface tension of the nucleation bubble interface between the false vacuum and the stable vacuum as a function of the temperature can be obtained, and the results are shown in Fig.\ref{Fig04} when the chemical potentials are taken as $\mu=306$ MeV and $\mu=309$ MeV. For both cases, with increasing of the temperature, the surface tension $\Sigma(T)$ starts to grow firstly and reaches a maximum at a certain temperature, then it will inflect and slope downwards. These nontrivial behaviors of $\Sigma(T)$ were also reported in a weak first-order phase transition \cite{Bessa:2008nw} and a strong first-order phase transition \cite{Zhou:2020bzk} by using the exact numerical method, but these nontrivial properties seem to be completely destroyed by the thin-wall approximation. As shown in the left panel of Fig.\ref{Fig02}, for a weak first-order phase transition, there exists a spinodal temperature $T_{c1}$ where a small barrier between the two minima in the potential will disappear. As a consequence, there is no bubble solution anymore as $T<T_{c1}$ and the surface tension decreases rapidly to zero when $T\rightarrow T_{c1}$. But for a strong first-order phase transition, according to a standard criterion to guarantee the existence of the stable bounce or soliton, it is indispensable for the potential of the order parameter $\sigma$ field to exhibit three distinct extrema. Hence, we can have a nontrivial solution to the equation of motion (\ref{eom}) for any temperature as $T<T_c$. The surface tension will then decline to a nonzero value as the decreasing of the temperature as depicted at the right panel of Fig.\ref{Fig04}. This is the main difference between a strong first-order phase transition and a weak one, this is the reason why we have separated the area of the first-order phase transition in phase diagram into two categories: a strong one and a weak one in the previous discussion. 

\begin{figure}[thbp]
\epsfxsize=9.0 cm \epsfysize=6.5cm
\epsfbox{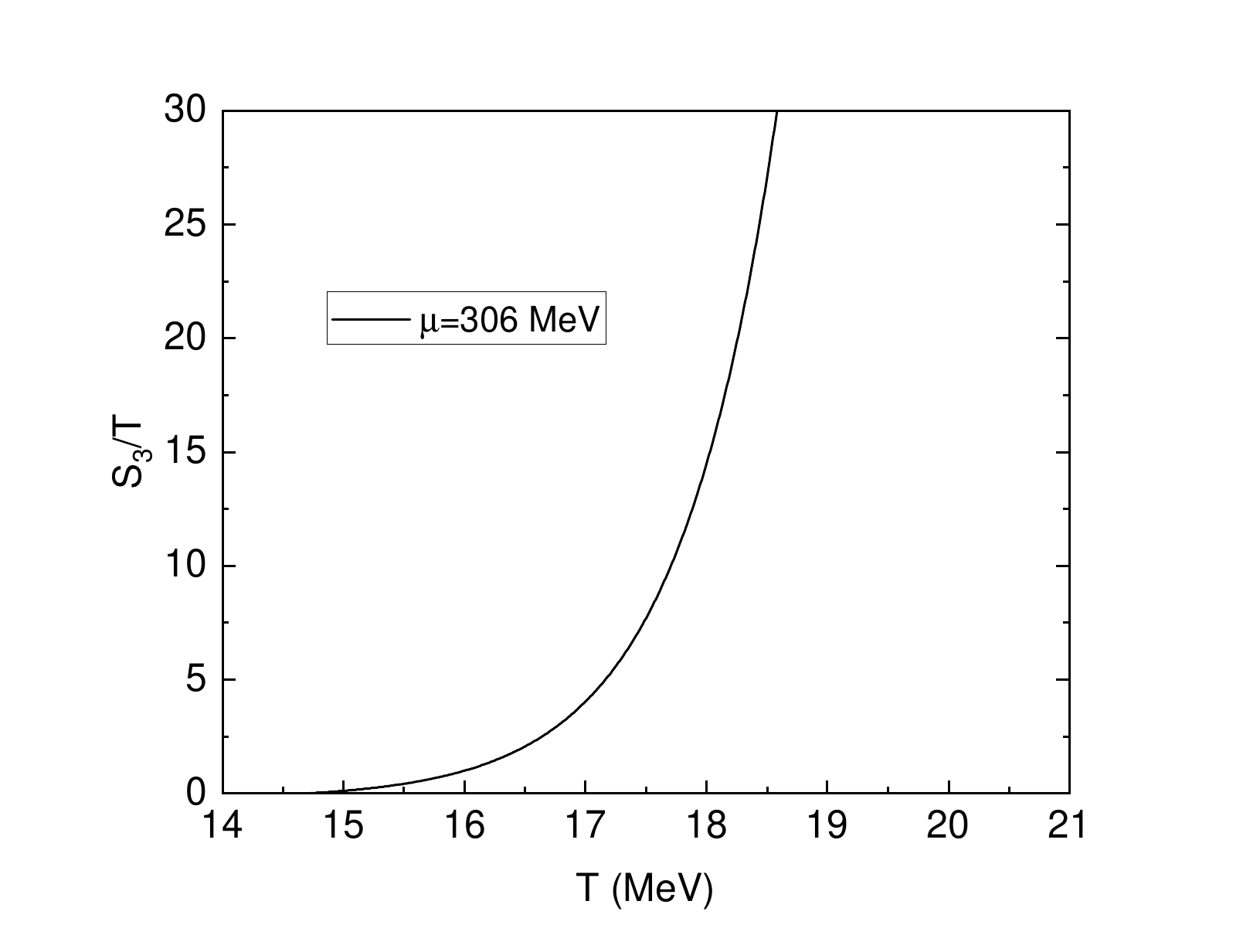}\hspace*{0.01cm} \epsfxsize=9.0 cm
\epsfysize=6.5cm \epsfbox{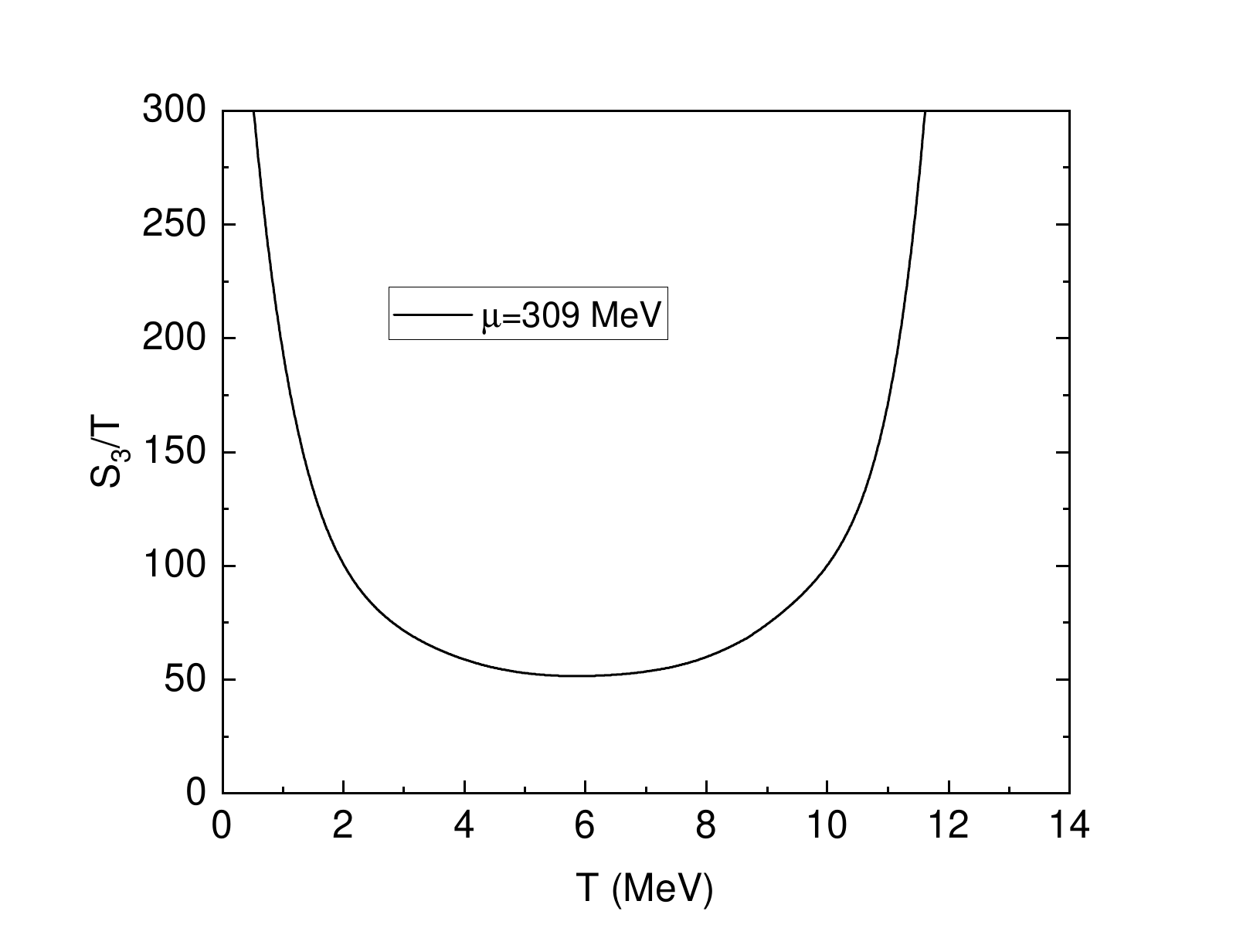}
 \caption{(Color online) (a) The saddle-point action as a function of temperature $T$ for $T\leq T_c$ at $\mu=306$ MeV. (b) The saddle-point action as a function of temperature $T$ for $T\leq T_c$ at $\mu=309$ MeV.}
\label{Fig05}
\end{figure}

We are now to determine the $S_3/T$ exponent in Eq.(\ref{nuclrate}), which is the saddle-point action evaluated on the bounce solution. Because the decay rate per unit volume is what we are interested, the argument of the exponential $S_3/T$ is more important in comparison with the prefactor $\Gamma$ if the $S_3/T$ is larger than the unity $1$. As shown in the following discussions, for the most studies in the present work, an estimate for the prefactor based on dimensional analysis is sufficient. To show the saddle-point action due to the appearance of the critical bubble and its crucial role played in the nucleation rate for the first-order phase transition, the $S_3/T$ exponent as a function of the temperature $T$ at different chemical potentials are plotted in Fig.\ref{Fig05}. As shown in the left panel in Fig.\ref{Fig05}, for a weak first-order phase transition as $\mu=306$ MeV, $S_3/T$ will start from zero when the temperature is at the spinodal temperature $T_{c1}$ because the barrier is to disappear and there is only a trivial solution for the field equation of the $\sigma$ field. Then it will climb up very quickly with the increase of the temperature and tend to diverge near the critical temperature $T_c$. By the exponential form of equation (\ref{nuclrate}), $\Gamma$ will be strongly suppressed by the saddle-point action and the system is likely to stay in the metastable vacuum for a relatively long time as long as $S_3/T>1$. Therefore, for a weak first-order phase transition, the system can be trapped in quark phase even when the temperature is below than the critical temperature $T_c$, until to the temperature such that $S_3/T\simeq 1$. After that, the exponential factor is unimportant and the probability of a false vacuum decay through the barrier is essentially enhanced by the thermodynamical fluctuation. From the left panel in Fig.\ref{Fig05}, for $\mu=306$ MeV, the remarkable temperature is about $T\simeq 16$ MeV when $S_3/T\simeq 1$, it is very close to the spinodal critical temperature $T_{c1}\simeq 14.7$ MeV. This indicates that the quark phase could survive safely up to the temperature nearby the low spinodal temperature $T_{c1}$ for a weak first-order phase transition.     

When fixing the chemical potential at $309$ MeV, the resulting plot of the $S_3/T$ exponent as a function of the temperature $T$ are shown in the right panel in Fig.\ref{Fig05}. In this case, the $S_3/T$ firstly decreases with the increase of the temperature, then it reaches a minimum point and restarts to grow up very quickly. As the temperature is close to the critical temperature $T_c$, it will become divergent. In comparison with the former case of a weak first-order phase transition, for a strong one, the saddle-point action as a function of temperature $T$ usually exhibits a non-monotonic behavior with increasing of the temperature, this can be taken as one of the salient properties of the strong first-order phase transition \cite{Zhou:2020bzk}. Besides the non-trivial property of the saddle-point action, there is another obvious difference from that of the first-order phase transitions. For $T<T_c$, no matter what the temperature we are taken, the $S_3/T$ is always larger than the unity $1$. Such that, for a strong first-order phase transition in the present model, it seems the conversion of quark phase to hadron phase is exponentially suppressed for $T\leq T_c$ and $\mu\geq \mu_c$ with $\mu_c\simeq 308$ MeV, and the system is likely to stay in quark phase rather than hadron phase, this could in turn induce various exotic structures of the phase of the strong-interaction matter in high density and low temperature due to the presence of the free quarks.      

\subsection{Bottom-up}

By ``bottom-up", we mean that the system is heating up from low energy to very high energy and the starting point is set as hadron phase when $T\simeq T_c$. With increasing of the temperature further, while $T>T_c$, the appearance of a bubble of the quark phase (the stable vacuum) inside the hadron phase (the false vacuum) is also treated as a natural consequence of the thermal fluctuations of the thermodynamical system sufficiently nearby the coexistence line in phase diagram. Therefore in the ``bottom-up" scenario when $T>T_c$, we should numerically solve the equation of motion in Eq.(\ref{eom}) with the specific boundary conditions as $\sigma\rightarrow \sigma_h$ as $r\rightarrow \infty$ and $d \sigma(0)/dr=0$ because the false vacuum is located at the hadron phase and the quark phase is now the stable state.

\begin{figure}[thbp]
\epsfxsize=9.0 cm \epsfysize=6.5cm
\epsfbox{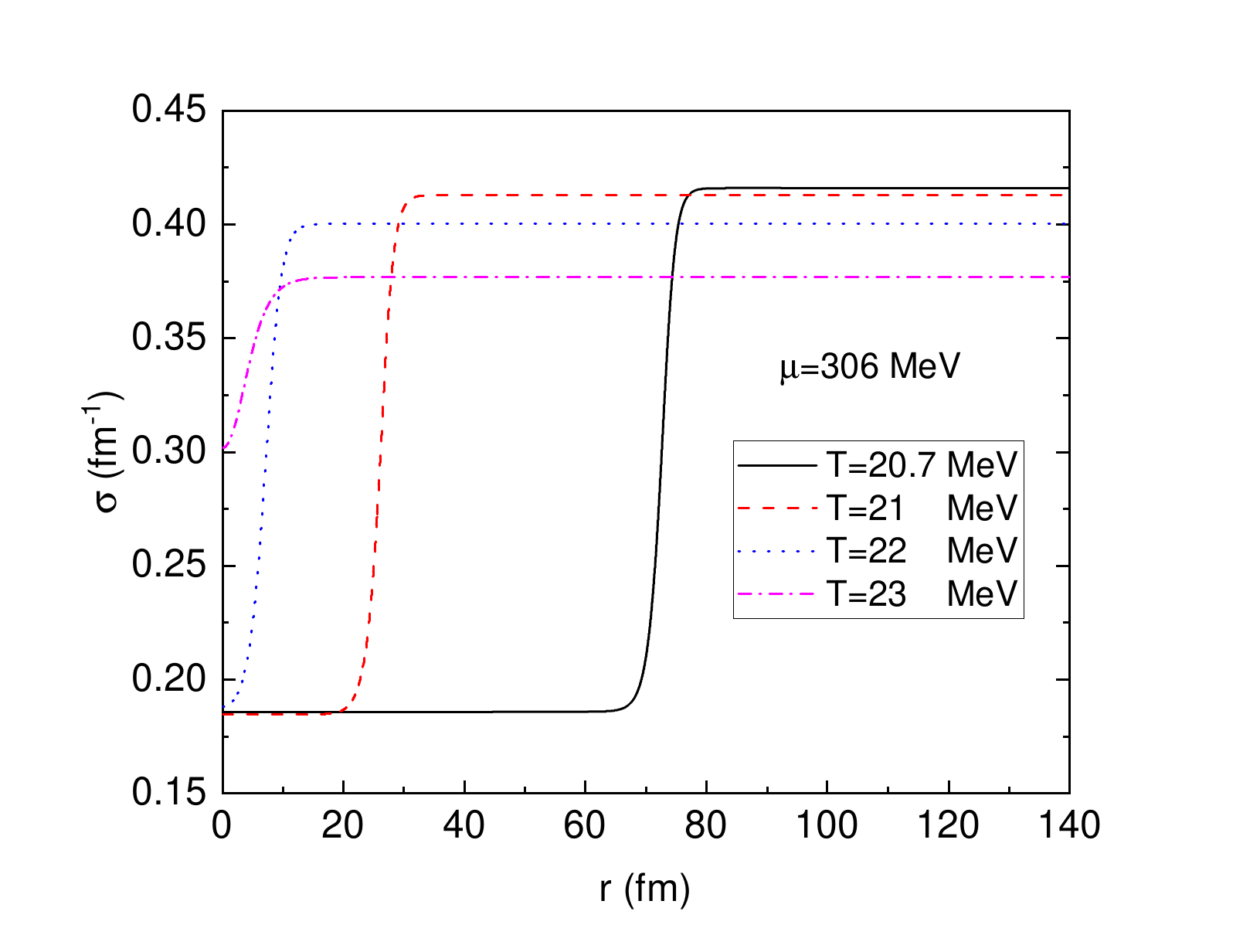}\hspace*{0.01cm} \epsfxsize=9.0 cm
\epsfysize=6.5cm \epsfbox{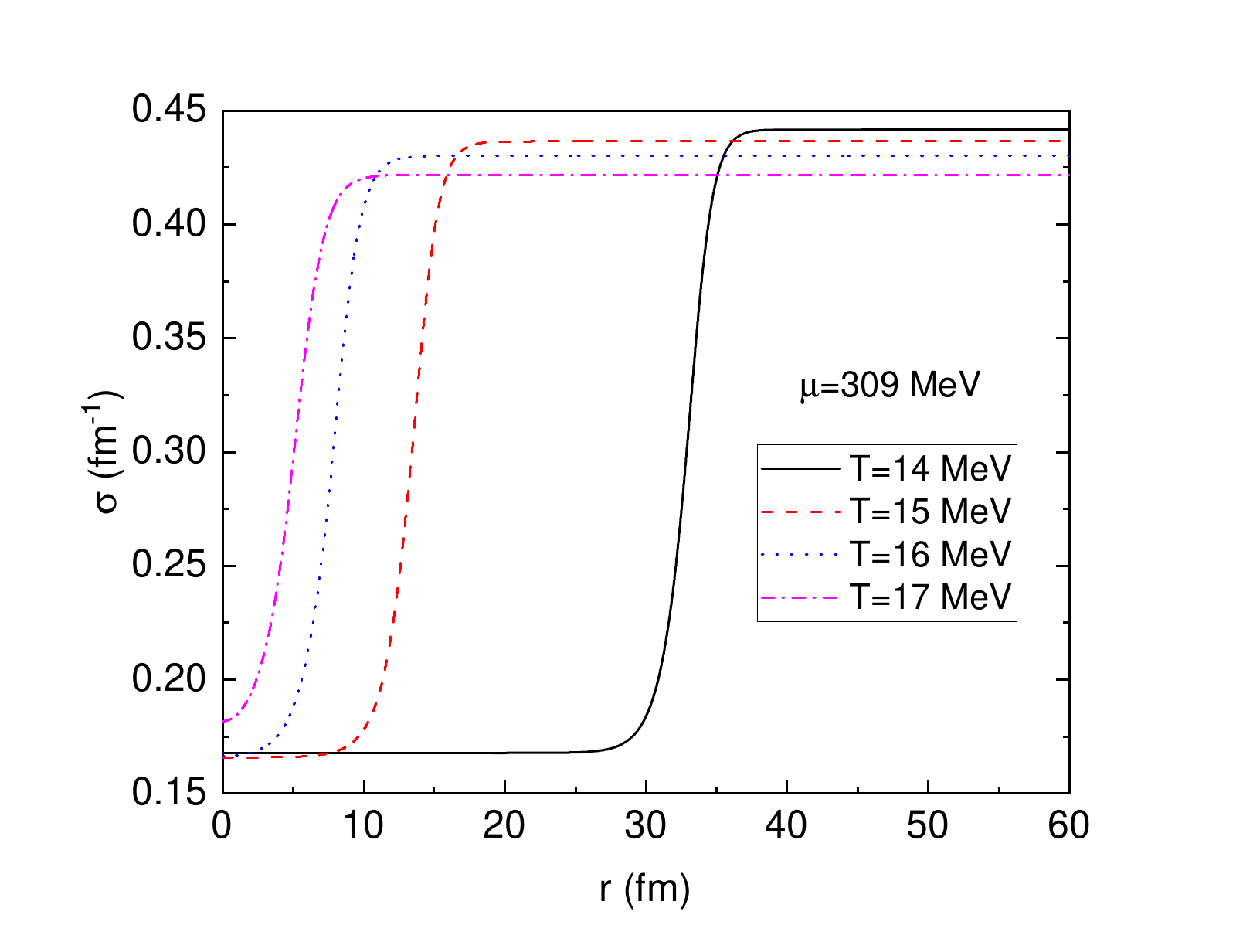}
 \caption{(Color online) (a) Critical bubble profiles for different temperatures when fixing the chemical potential at $\mu=306$ MeV for $T>T_c$. From right to left, the curves correspond to $T=20.7$, $21$, $22$ and $23$ MeV. (b) Critical bubble profiles for different temperatures when fixing the chemical potential at $\mu=309$ MeV for $T>T_c$. From right to left, the curves correspond to $T=14$, $15$, $16$ and $17$ MeV.}
\label{Fig06}
\end{figure}

In what follows, we begin by showing the critical bubble profiles obtained from the exact numerical solution of Eq.(\ref{eom}) with the boundary conditions $\sigma(\infty)=\sigma_h$ and $\sigma'(0)=0$ when fixing chemical potentials at $\mu=306$ MeV and $\mu=309$ MeV in Fig.\ref{Fig06}. This is unlike the results presented in the ``top-down" scenario when $T<T_c$, in which there have two different first-order phase transitions. However, for $T>T_c$, the results for both panels in Fig.\ref{Fig06} exhibit some quite similar features. For both cases, with the increase of the temperature from the critical temperature to the up spinodal line at $T=T_{c2}$, the typical size of the bubble which can be approximately estimated by the maximal value of the quantity $\left | \sigma'(r) \right |$ will decrease rapidly to zero because the barrier between the two minima in the potential disappears and there is no stable bubble solution anymore as $T\geq T_{c2}$. Furthermore, the structures of the critical bubbles also share alike properties. When the temperature is close to the critical temperature $T_c$, the critical bubble has an obvious ``core" structure with $\sigma \simeq \sigma_l$ separated by a relatively thin wall from the region $\sigma \simeq \sigma_h$. On the other side, when the temperature comes up to another critical point at $T\simeq T_{c2}$, the critical bubble usually becomes ``coreless" structure due to the fact that the thickness of the critical bubble has the same order as the radius and the field at the original point $\sigma(r=0)$ departs from its true vacuum $\sigma_l$ largely. At last, the curves in Fig.\ref{Fig06} can be explained qualitatively according to the ``overshoot-undershoot" argument given by Coleman. When the temperature is very close to the critical temperature $T_c$, the potential has two degenerate vacua, and the damping force is neglectful small, the field at escape point $\sigma(r=0)$ starts at the top of the effective potential around $\sigma\simeq \sigma_l$. On the contrary, with the temperature rising up, two degenerate vacua get decoupled and the damping force takes effect, the field $\sigma(r=0)$ will deviate from its vacuum value dramatically, especially as $T\rightarrow T_{c2}$. In other words, the thin-wall approximation is not expected to be valid, any further extension of the thin-wall approximation to higher temperatures should be checked very carefully particularly when the temperature is close to the up spinodal line.   

\begin{figure}[thbp]
\epsfxsize=9.0 cm \epsfysize=6.5cm
\epsfbox{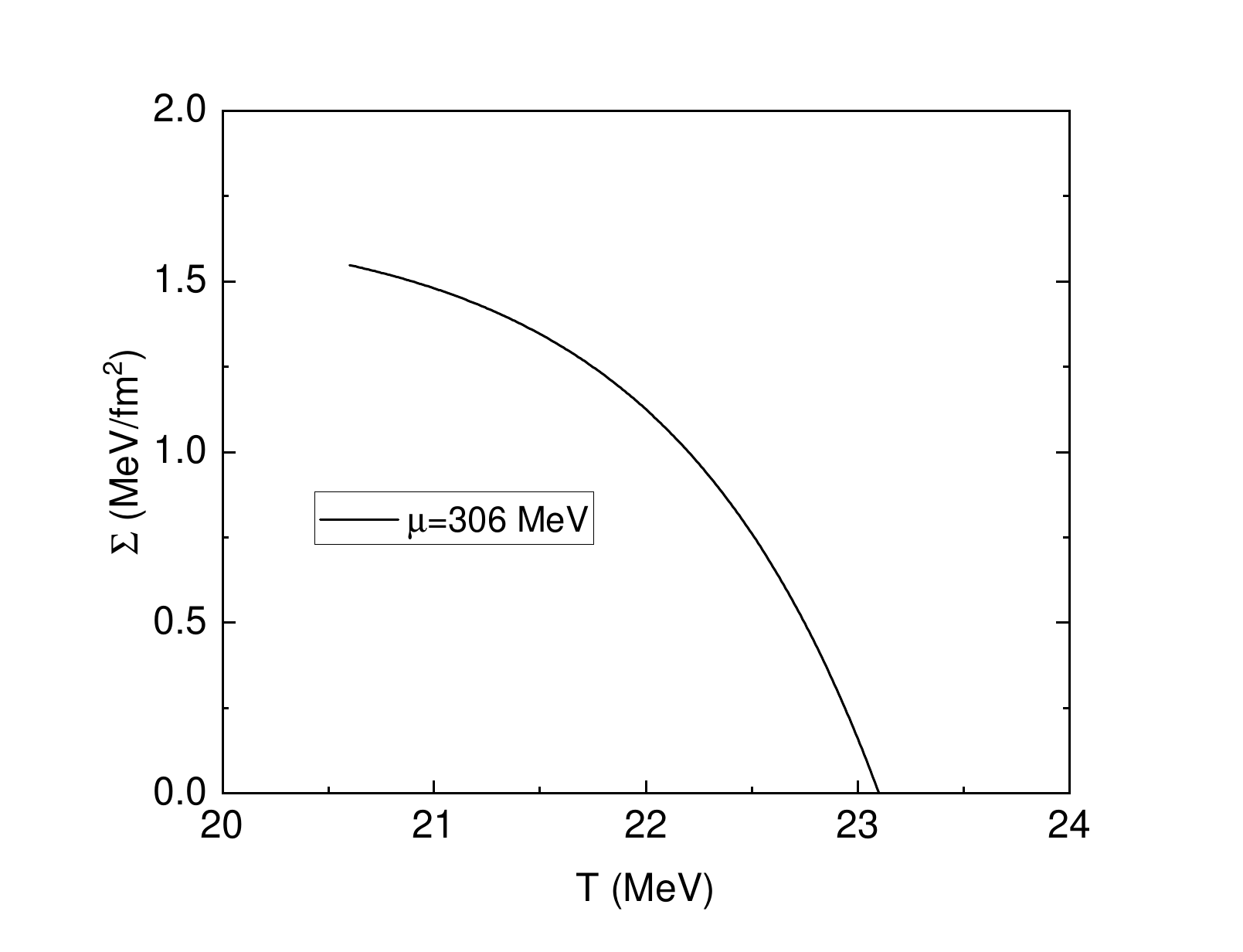}\hspace*{0.01cm} \epsfxsize=9.0 cm
\epsfysize=6.5cm \epsfbox{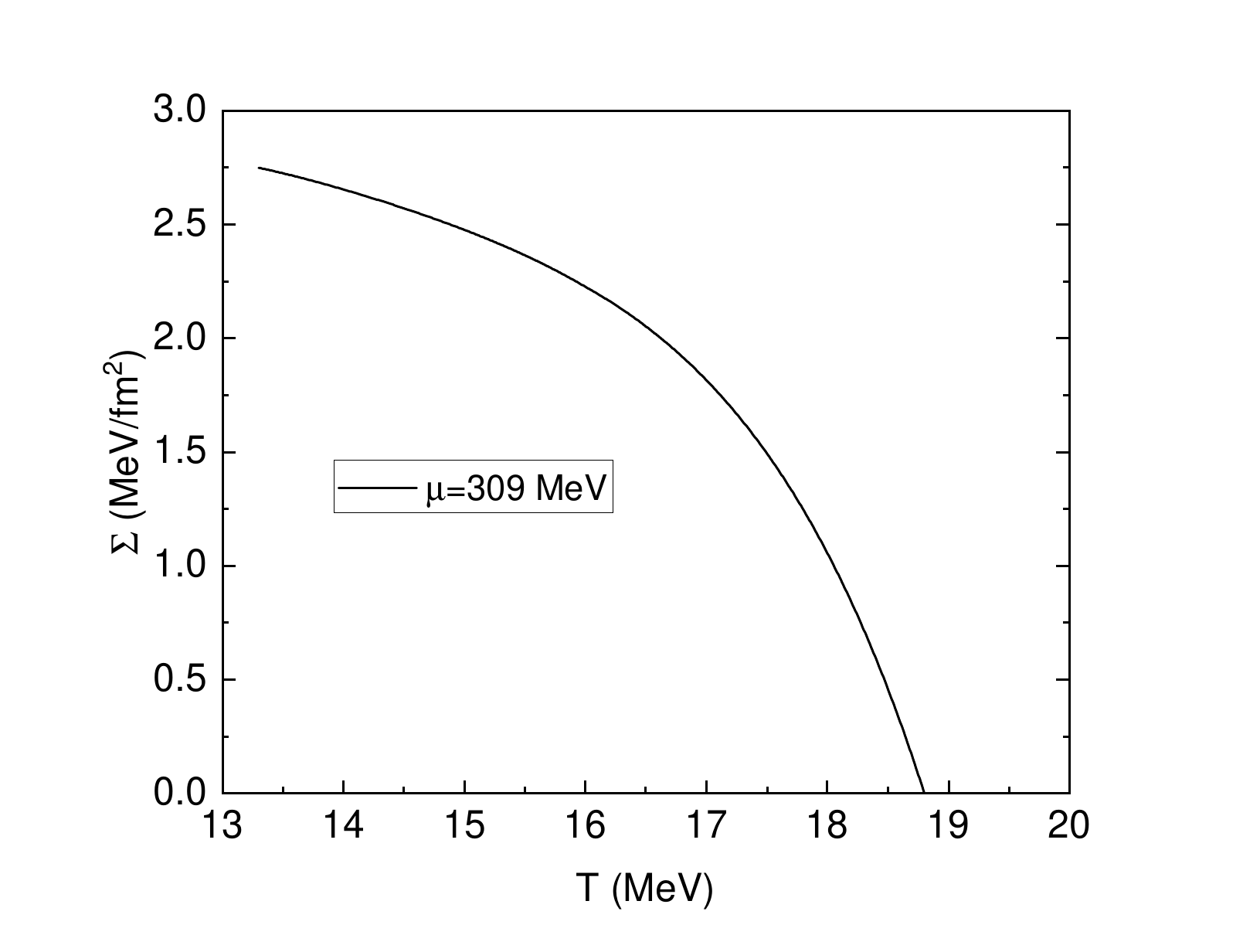}
 \caption{(Color online) (a) Surface tension as a function of temperature $T$ for $T\geq T_c$ at $\mu=306$ MeV. (b) Surface tension as a function of temperature $T$ for $T\geq T_c$ at $\mu=309$ MeV.}
\label{Fig07}
\end{figure}

From the definition of the surface tension in Eq.(\ref{surfacet}), we plot the temperature dependence of the surface tensions for $T\geq T_c$ when fixing the chemical potential at $\mu=306$ MeV and $\mu=309$ MeV in Fig.\ref{Fig07}. For $\mu=306$ MeV, the values of the surface tension are between about $1.6$ MeV and $0$ MeV, while for a relatively larger chemical potential $\mu=309$ MeV, they are among $\sim2.8$ MeV to zero. Then, we can find the biggest values of the surface tension occur near the critical line since this domain is characterized by large barriers and a small energy difference between the true and false vacua. Besides, the result in Fig.\ref{Fig07} also implies that the surface tension nearby the critical line at $T\simeq T_c$ will increase accordingly with the increase of the chemical potential. Moreover, for both cases, the surface tensions will continuously decline to zero as long as the temperature approaches to the up spinodal line at $T\simeq T_{c2}$. Therefore, in the bottom-up scenario, $\Sigma(T)$ is a monotonically decreasing function of $T$, whereas in the top-down scenario it is a non-monotonic function and has a nontrivial behavior. Sometimes, the nontrivial evolution of $\Sigma(T)$ suggests that the temperature dependent surface tension has a maximum value at a specific temperature and it can be taken as a limit to the applicability of the thin-wall approximation \cite{Zhou:2020bzk,Bessa:2008nw}. Thus, for the bottom-up scenario, we need to develop an alternative method to estimate a scope in which the thin-wall approximation is valid.

\begin{figure}[thbp]
\epsfxsize=9.0 cm \epsfysize=6.5cm
\epsfbox{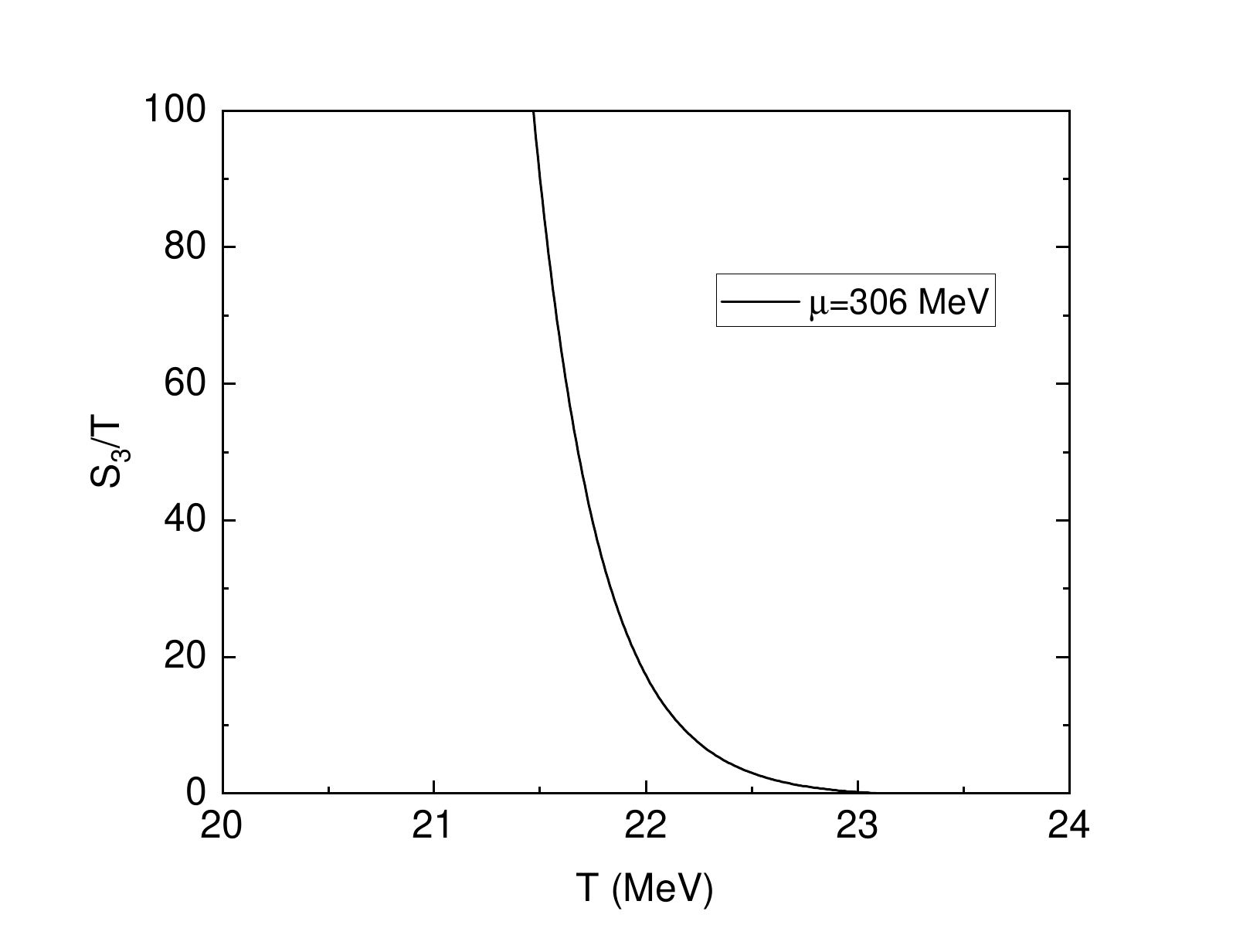}\hspace*{0.01cm} \epsfxsize=9.0 cm
\epsfysize=6.5cm \epsfbox{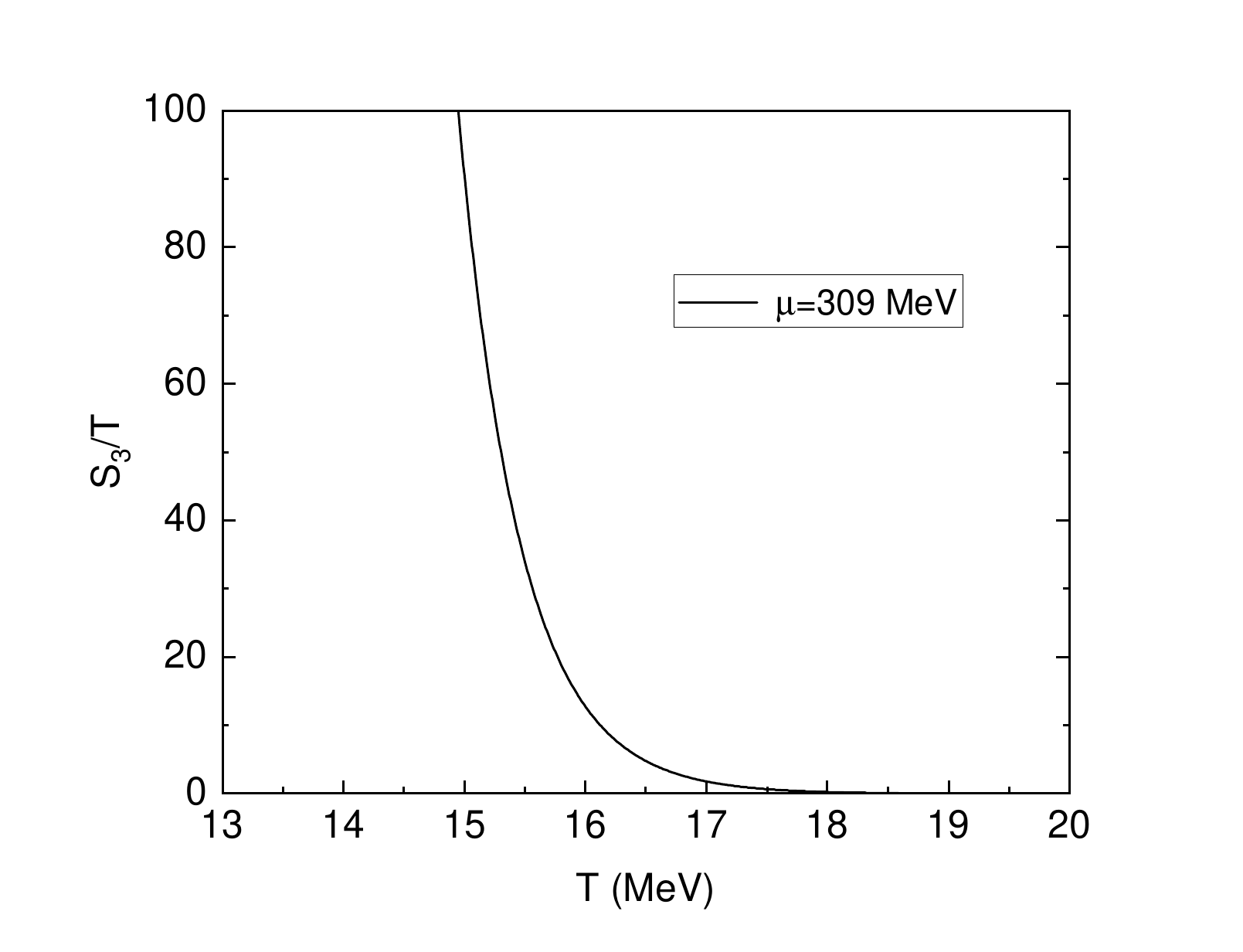}
\caption{(Color online) (a) The saddle-point action as a function of temperature $T$ for $T\geq T_c$ at $\mu=306$ MeV. (b) The saddle-point action as a function of temperature $T$ for $T\geq T_c$ at $\mu=309$ MeV.}
\label{Fig08}
\end{figure}
  
To study the dynamics of a first-order phase transition, the last important quantity to be evaluated is the saddle-point action $S_3/T$ due to the activation of a nucleation bubble, which is an essential ingredient for the nucleation rate per unit time per unit volume in Eq.(\ref{nuclrate}). In Fig.\ref{Fig08}, $S_3/T$ is plotted as a function of temperature $T$ for $T\geq T_c$ when fixing the chemical potentials at $\mu=306$ MeV and $\mu=309$ MeV. From this figure, as the temperature approaches from above to the critical temperature $T_c$, $S_3/T$ will increase every quickly and diverges near the critical temperature $T_c$. In the opposite direction, when the temperature is close to the spinodal critical temperature $T_{c2}$, $S_3/T$ will decrease rapidly to zero, during this procedure, what we are interested in is the moment when $S_3/T$ is about a unity one because if $S_3/T>1$ the nucleation rate $\Gamma$ will be strongly suppressed by the exponential factor and the system is likely to stay in the false vacuum for a very long time. From the left panel of Fig.\ref{Fig08}, for $\mu=306$ MeV, when the temperature is about $22.8$ MeV, $S_3/T\simeq 1$, so that the system is likely to remain in hadron phase until the temperature is less than $22.8$ MeV. For $\mu=309$ MeV, this specific temperature is approximately $18.4$ MeV when $S_3/T\simeq 1$, this indicates that hadron phase could also maintain its existence as long as the temperature is below $18.4$ MeV. Since these two specific temperatures are very close to their spinodal critical temperatures $T_{c2}$, to a rough estimation, we can simply taken the spinodal critical line in phase diagram as a phase boundary for the stable existence of the false vacuum in the first-order phase transition.     

\section{Conclusion}

In this work we have computed the effective potential for two-flavor quark-meson model at finite temperature and density in the presence of a fermionic vacuum term. Having the in-medium effective potential, the phase diagram together with the critical end point has been given and the up and low spinodal lines have been calculated explicitly for the first-order hadron quark phase transition. For the low spinodal line, the first-order phase transition can further be divided into a strong and weak ones in phase diagram when the temperature is below the critical coexistence line. The critical chemical potential is taken as $\mu\simeq 308$ MeV as the low spinodal line terminates at this point. So that as $T<T_c$, for $\mu<308$ MeV, it is a weak first-order phase transition, but for $\mu>308$ MeV, it belongs to a strong first-order phase transition.  

Provided by the temperature-dependent effective potential, the problem of homogeneous nucleation of bubble in a first-order phase transition can be investigated accordingly. For convenience, we have separated our discussions into two scenarios: the top-down case and the bottom-up case. By ``top-down”, we consider the quark phase as a metastable (false) vacuum and the hadron phase as a stable (true) vacuum when $T<T_c$. On the contrary, by ``bottom -up", we mean that the false vacuum is the hadron phase whereas the true vacuum is the quark phase as $T>T_c$. Then for the former case, the boundary condition at infinite radius is $\sigma=\sigma_h$, whereas it is $\sigma=\sigma_l$ for the latter case. With these specific boundary conditions, a saddle point solution of the field equation has been solved and the exact bubble profiles were obtained. Usually, when the temperature is close to the critical temperature $T_c$, the bubble profile shows a ``core" structure with the sigma field at true vacuum $\sigma \simeq \sigma_{T}$ separated by a relatively thin wall from the false vacuum at $\sigma \simeq \sigma_{FV}$. However, when the temperature approaches to the spinodal critical temperature, the bubble profile exhibits a coreless structure since now the thickness of the critical bubble has the order of the radius and the $\sigma$ field inside the bubble departs from its true vacuum value significantly.       

\begin{figure}
\includegraphics[scale=0.36]{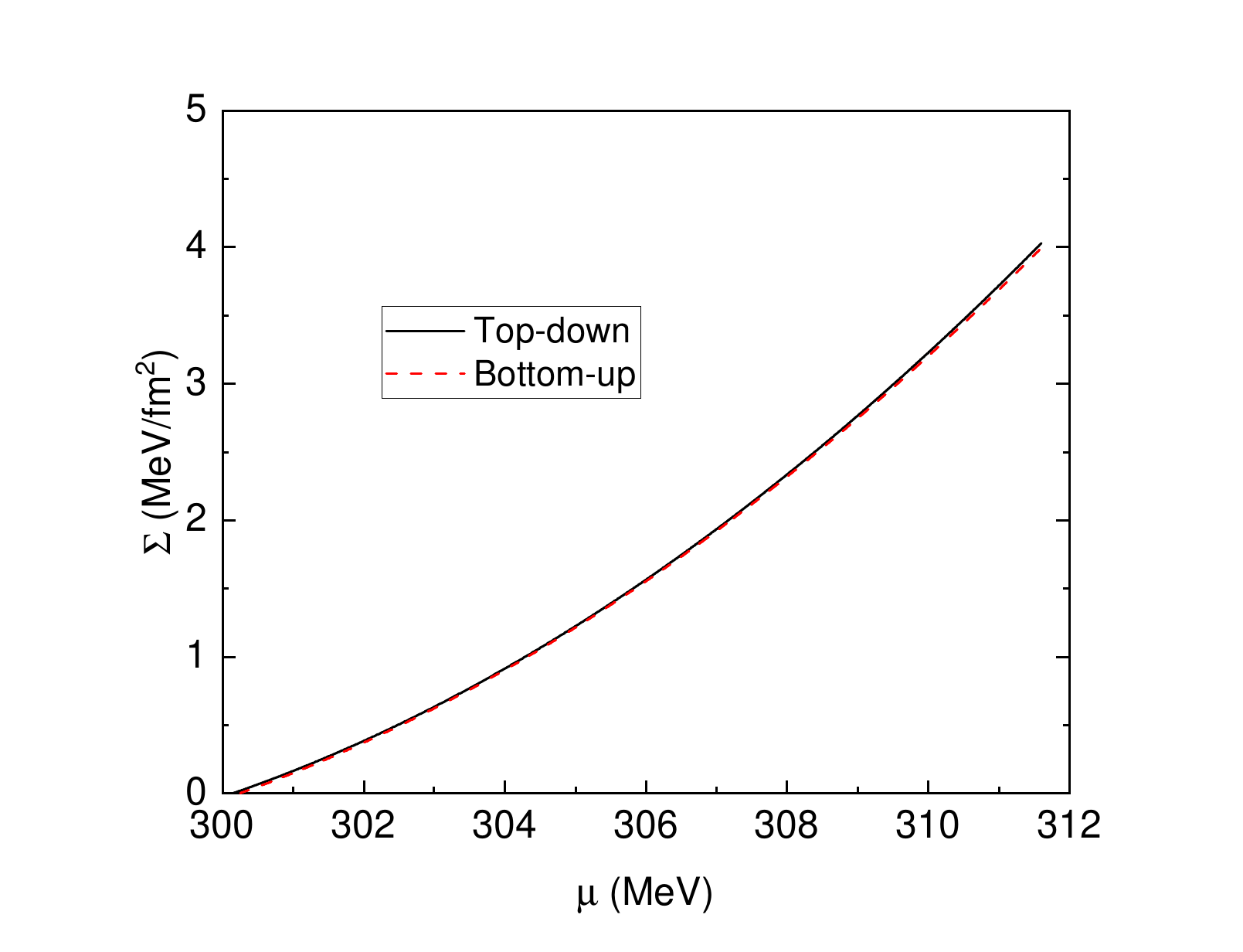}
\caption{\label{Fig09} (Color online) Surface tension as a function of a quark chemical potential when $T\simeq T_c$. The solid line is for the ``top-down" scenario and the dashed line is for the ``bottom-up" scenario.}
\end{figure} 

The calculation for the surface tension between a quark phase and hadron phase were also presented in these two different scenarios. In the ``top-down" context, the surface tension first increases to a maximum value and then decrease with the increase of the temperature. The top of the surface tension could be taken as a limit on the reliability of the thin-wall approximation because the bubble profile at this point hints of a largest distortion of that of the thin-wall approximation. On the other side, for the ``bottom-up" context, the surface tension demonstrates a monotonic property with the growth of the temperature, it will descend continually from its top value at the critical temperature $T_c$ to zero as $T\rightarrow T_{c2}$. As we known, the surface tension plays an important role in the field of nuclear physics and astrophysics and has attracted much attention recently. To provide a comprehensive consultation of the relevant research, the surface tension along the critical coexistence line has been laid out in Fig.\ref{Fig09}, we believe the remaining value of the surface tension in the domain of the first-order phase transition can be easily extracted and estimated by using the method in this work. Consequently, the present model predicts a surface tension of $\Sigma\sim 0-4$ $\mathrm{MeV/fm^2}$, our results are very close to the ones recently found for the same model in the thin-wall approximation in Ref \cite{Wan-Ping:2019ifh} since the thin-wall approximation is a reliable tool as $T\simeq T_c$. Note that most effective models also give a predict for a small values as $\Sigma\leq 30$ $\mathrm{MeV/fm^2}$, such as the MIT bag model\cite{Oertel:2008wr}, NJL model\cite{Garcia:2013eaa,Ke:2013wga}, three-flavor PQM model\cite{Mintz:2012mz}, the chiral nucleon-meson model\cite{Fraga:2018cvr} and the Friedberg-Lee model \cite{Zhou:2020bzk}. Such a small value of the surface tension would lead to a quark-matter formation during core-collapse supernova explosion and favors a mixed phase in the cores of compact stars, so that this reasonably small surface tension could provide an observable signal of the first-order phase transition within compact stars and play an important role in astrophysics.  

It is interesting to point out that most studies in literature on the subject of the surface tension are based on the quark model. Such as the MIT model, the NJL model, the QM model, or their modernized versions, the PNJL and PQM model. Taking the QM model as a typical example, since there only has the degree of freedom of the quarks, although the model can be successfully used to study the quark phase, it can not be used to describe the properties of nuclear matter directly. Of course, based on the picture of the non-topological soliton, we can in principle reconstruct and get the correct degree of freedom of the nucleons in hadron phase as done in Refs.\cite{Mao:2013qu,Jin:2015goa}. But such a consideration will lose the knowledge of the entire potential which is a necessary condition to find the bounce solution during the hadron quark phase transition. To compensate this problem, it is indispensable for studying the first-order phase transition based on other effective models which can give a successful description of the properties of nuclear matter at saturation density, the nuclear liquid-gas phase transition or even in the region of a hadron quark phase transition. Fortunately, the chiral nucleon-meson model \cite{Floerchinger:2012xd}\cite{Drews:2013hha} fulfils the condition and the surface tension has been calculated and obtained in Ref.\cite{Fraga:2018cvr}. From this work, a small value of the surface tension is also reported and the results obtained in there are similar to our results as $T\rightarrow T_c$. Furthermore, besides the hadron quark phase transition, the first-order nuclear liquid-gas phase transition and the surface tension have been also investigated in Ref.\cite{Fraga:2018cvr}. Therefore, by combined their results with our present study, we can give out a comprehensive picture of a hadron quark phase transition from the vacuum to the quark phase totally. 

For a weak first-order phase transition, our results show a rapid decrease with temperature in the saddle-point action of critical bubbles from infinity at the critical temperature to zero at the spinodal critical temperature. This implies that we can always have the moment at which $S_3/T\simeq 1$ for the weak first-order phase transition. However, for a strong first-order phase transition, the saddle-point action of critical bubbles shows some quite different characterizations. When the temperature increases, it first decreases to a minimum value, then it rises up and diverges as the temperature is close to the critical temperature $T_c$. This non-monotonic behavior of the $S_3/T$ with the increase of the temperature also reported in a previous study based on the Friedberg-Lee model \cite{Zhou:2020bzk}, thus it can be taken as a salient feature of a strong first-order phase transition in comparison with the weak ones. Another interesting character of the $S_3/T$ is that the saddle-point action never gets chance to reach the unity $1$ for a strong first-order phase transition. The result can be roughly interpreted with the former study in Fig.6 in Ref.\cite{Zhou:2020bzk}, where the evolution of the $S_3/T$ as a function of chemical potential shows that for a fixed temperature the saddle-point action will increase and go across unity $1$ quickly when the chemical potential rises up to $\sim231$ MeV. Hence the $S_3/T$ is believed to satisfy the condition $S_3/T>1$ in the present study since we have a lower temperature and larger chemical potential. Given the exponential dependence of $\Gamma$ on $S_3/T$, the decay of the false vacuum is to be exponentially suppressed and the system is likely to stay in the metastable state for a relatively long time when $S_3/T>1$. So that the false vacuum could survive and exist as a metastable state as long as the temperature lies between the up and low spinodal lines, only if the temperature is close to the spinodal critical line, the $S_3/T$ will descend and go through the unity $1$. More specifically, a ``conventional" hadron matter among the critical coexistence line and the low spinodal line could be potentially treated as a quark matter, then the exotic structures of the strong-interaction matter predicted by theoretical calculations should be reconsidered and revised accordingly, such as the quarkyonic matter \cite{McLerran:2007qj,Fukushima:2013rx}, pion superfluidity \cite{He:2005nk,Adhikari:2018cea}, color superconductors \cite{Rajagopal:2000wf,Alford:2007xm} and the inhomogeneous chiral condensates along the first-order phase transition line \cite{Buballa:2014tba}. On the contrary, based on QCD theoretical calculations, besides the CEP already discussed in above, there is also a possibility that the first-order phase boundary ends at another critical point in the lower-temperature and higher-density region according to the so called the quark-hadron continuity \cite{Ma:2019ery}\cite{Baym:2017whm}. Below such a CEP, the order of QCD phase transition is crossover because the cold dense QCD matter with three degenerate flavours may have no clear border between superfluid nuclear matter and superconducting quark matter. This means that our study should be constrained on a very narrow region in the first-order phase transition, even there is no first-order phase transition by some possibility. Therefore, the results about the very low temperature should be modified accordingly. 

Of course,in the present study,we have only simply considered the mean-field approximation and the role of the important thermodynamic fluctuations has been ignored completely. Any attempts to go beyond the mean-field approximation are worthy to exploring in order to identify their effects in the studies on phase diagram, the order of phase transition and particularly the dynamics of the first-order phase transition, among these methods, a functional renormalization group \cite{Schaefer:2004en,Fu:2022gou,Chen:2023cjt} and the Cornwall–Jackiw–Tomboulis (CJT) formalism based on the two-particle irreducible (2PI) action \cite{Cornwall:1974vz,Petropoulos:2004bt,Mao:2013gva} are popular approaches. Such an extension is straightforward but technically complicated. Moreover, besides thermal fluctuations, quantum fluctuations have been simply ignored in the present study. According the homogenous nucleation theory, when the temperature is very low, as $R\leq \beta=1/T$, the argument suggested by Linde \cite{Linde:1981zj} is not going to satisfy and the tunneling of the false vacuum will be induced by quantum fluctuations. At the moment, we have to calculate the tunneling rate for $O(4)$ bounce in Eq.(\ref{Euaction}) rather than the three-dimensional action in Eq.(\ref{Euaction2}). Therefore, all the calculations included the low temperature limit are considered in the light of an assumption that thermal fluctuations dominate quantum fluctuations. Since the thermal fluctuations are suppressed when the temperature is very low, the quantum fluctuations are believed to take effect and play a dominant role. Here, how to incorporate the quantumly-induced tunneling rate as well as the thermally-induced tunneling rate in the quark-meson model simultaneously are left for future considerations. We believe all these studies in both directions will make us getting closer to the real QCD world.

\begin{acknowledgments}
We thank Ang Li, Jinshuang Jin, Shijun Mao, Ken D. Olum and Xinjian Wen for valuable comments and discussions. This work is supported in part by National Natural Science Foundation of China (NSFC) under No.11675048.
\end{acknowledgments}


\begin{thebibliography}{199}

\bibitem{Yagi:2005yb}
K.~Yagi, T.~Hatsuda and Y.~Miake,
Camb. Monogr. Part. Phys. Nucl. Phys. Cosmol. \textbf{23}, 1-446 (2005)

\bibitem{Fukushima:2010bq}
K.~Fukushima and T.~Hatsuda,
Rept. Prog. Phys. \textbf{74}, 014001 (2011)
doi:10.1088/0034-4885/74/1/014001
[arXiv:1005.4814 [hep-ph]].

\bibitem{Braun-Munzinger:2015hba}
P.~Braun-Munzinger, V.~Koch, T.~Sch\"afer and J.~Stachel,
Phys. Rept. \textbf{621}, 76-126 (2016)
doi:10.1016/j.physrep.2015.12.003
[arXiv:1510.00442 [nucl-th]].

\bibitem{Luo:2017faz}
X.~Luo and N.~Xu,
Nucl. Sci. Tech. \textbf{28}, no.8, 112 (2017)
doi:10.1007/s41365-017-0257-0
[arXiv:1701.02105 [nucl-ex]].

\bibitem{Pandav:2022xxx}
A.~Pandav, D.~Mallick and B.~Mohanty,
Prog. Part. Nucl. Phys. \textbf{125}, 103960 (2022)
doi:10.1016/j.ppnp.2022.103960
[arXiv:2203.07817 [nucl-ex]].

\bibitem{Aarts:2023vsf}
G.~Aarts, J.~Aichelin, C.~Allton, A.~Athenodorou, D.~Bachtis, C.~Bonanno, N.~Brambilla, E.~Bratkovskaya, M.~Bruno and M.~Caselle, \textit{et al.}
Prog. Part. Nucl. Phys. \textbf{133}, 104070 (2023)
doi:10.1016/j.ppnp.2023.104070
[arXiv:2301.04382 [hep-lat]].

\bibitem{Nambu:1961tp}
  Y.~Nambu and G.~Jona-Lasinio,
  Phys.\ Rev.\  {\bf 122}, 345 (1961);
    
\bibitem{Nambu:1961fr}
  Y.~Nambu and G.~Jona-Lasinio,
  Phys.\ Rev.\  {\bf 124}, 246 (1961).

 
\bibitem{Gell-Mann:1960mvl}
M.~Gell-Mann and M.~Levy,
Nuovo Cim. \textbf{16}, 705 (1960)
doi:10.1007/BF02859738
  
\bibitem{Costa:2010zw}
  P.~Costa, M.~C.~Ruivo, C.~A.~de Sousa and H.~Hansen,
  Symmetry {\bf 2}, 1338 (2010), and references therein.
 
\bibitem{Schaefer:2007pw}
  B.~J.~Schaefer, J.~M.~Pawlowski and J.~Wambach,
  Phys.\ Rev.\  D {\bf 76}, 074023 (2007).
 
\bibitem{Mao:2009aq}
H.~Mao, J.~Jin and M.~Huang,
J. Phys. G \textbf{37}, 035001 (2010)
doi:10.1088/0954-3899/37/3/035001
[arXiv:0906.1324 [hep-ph]].

\bibitem{Schaefer:2009ui}
B.~J.~Schaefer, M.~Wagner and J.~Wambach,
Phys. Rev. D \textbf{81}, 074013 (2010)
doi:10.1103/PhysRevD.81.074013
[arXiv:0910.5628 [hep-ph]].
  
\bibitem{Fukushima:2017csk}
K.~Fukushima and V.~Skokov,
Prog. Part. Nucl. Phys. \textbf{96}, 154-199 (2017)
doi:10.1016/j.ppnp.2017.05.002
[arXiv:1705.00718 [hep-ph]].

\bibitem{LIGOScientific:2016aoc}
B.~P.~Abbott \textit{et al.} [LIGO Scientific and Virgo],
Phys. Rev. Lett. \textbf{116}, no.6, 061102 (2016)
doi:10.1103/PhysRevLett.116.061102
[arXiv:1602.03837 [gr-qc]].

\bibitem{Caprini:2019egz}
C.~Caprini, M.~Chala, G.~C.~Dorsch, M.~Hindmarsh, S.~J.~Huber, T.~Konstandin, J.~Kozaczuk, G.~Nardini, J.~M.~No and K.~Rummukainen, \textit{et al.}
JCAP \textbf{03}, 024 (2020)
doi:10.1088/1475-7516/2020/03/024
[arXiv:1910.13125 [astro-ph.CO]].

\bibitem{Hindmarsh:2020hop}
M.~B.~Hindmarsh, M.~L\"uben, J.~Lumma and M.~Pauly,
SciPost Phys. Lect. Notes \textbf{24}, 1 (2021)
doi:10.21468/SciPostPhysLectNotes.24
[arXiv:2008.09136 [astro-ph.CO]].

\bibitem{Croon:2023zay}
D.~Croon,
[arXiv:2307.00068 [hep-ph]].

\bibitem{Paschalidis:2017qmb}
V.~Paschalidis, K.~Yagi, D.~Alvarez-Castillo, D.~B.~Blaschke and A.~Sedrakian,
Phys. Rev. D \textbf{97}, no.8, 084038 (2018)
doi:10.1103/PhysRevD.97.084038
[arXiv:1712.00451 [astro-ph.HE]].

\bibitem{Sieniawska:2018zzj}
M.~Sieniawska, W.~Turczanski, M.~Bejger and J.~L.~Zdunik,
Astron. Astrophys. \textbf{622}, A174 (2019)
doi:10.1051/0004-6361/201833969
[arXiv:1807.11581 [astro-ph.HE]].

\bibitem{Han:2018mtj}
S.~Han and A.~W.~Steiner,
Phys. Rev. D \textbf{99}, no.8, 083014 (2019)
doi:10.1103/PhysRevD.99.083014
[arXiv:1810.10967 [nucl-th]].

\bibitem{Xia:2019pnq}
C.~J.~Xia, T.~Maruyama, N.~Yasutake and T.~Tatsumi,
Phys. Rev. D \textbf{99}, no.10, 103017 (2019)
doi:10.1103/PhysRevD.99.103017
[arXiv:1902.08766 [hep-ph]].

\bibitem{Baiotti:2019sew}
L.~Baiotti,
Prog. Part. Nucl. Phys. \textbf{109}, 103714 (2019)
doi:10.1016/j.ppnp.2019.103714
[arXiv:1907.08534 [astro-ph.HE]].

\bibitem{Blacker:2020nlq}
S.~Blacker, N.~U.~F.~Bastian, A.~Bauswein, D.~B.~Blaschke, T.~Fischer, M.~Oertel, T.~Soultanis and S.~Typel,
Phys. Rev. D \textbf{102}, no.12, 123023 (2020)
doi:10.1103/PhysRevD.102.123023
[arXiv:2006.03789 [astro-ph.HE]].

\bibitem{Cao:2018tzm}
G.~Cao and S.~Lin,
[arXiv:1810.00528 [nucl-th]].

\bibitem{Csernai:1992tj}
L.~P.~Csernai and J.~I.~Kapusta,
Phys. Rev. D \textbf{46}, 1379-1390 (1992)
doi:10.1103/PhysRevD.46.1379

\bibitem{Kapusta:2006pm}
J.~I.~Kapusta and C.~Gale,
Cambridge University Press, 2011,
ISBN 978-0-521-17322-3, 978-0-521-82082-0, 978-0-511-22280-1
doi:10.1017/CBO9780511535130

\bibitem{Langer:1967ax}
J.~S.~Langer,
Annals Phys. \textbf{41}, 108-157 (1967)
doi:10.1016/0003-4916(67)90200-X

\bibitem{Langer:1969bc}
J.~S.~Langer,
Annals Phys. \textbf{54}, 258-275 (1969)
doi:10.1016/0003-4916(69)90153-5

\bibitem{Coleman:1977py}
S.~R.~Coleman,
Phys. Rev. D \textbf{15}, 2929-2936 (1977)
[erratum: Phys. Rev. D \textbf{16}, 1248 (1977)]
doi:10.1103/PhysRevD.16.1248

\bibitem{Callan:1977pt}
C.~G.~Callan, Jr. and S.~R.~Coleman,
Phys. Rev. D \textbf{16}, 1762-1768 (1977)
doi:10.1103/PhysRevD.16.1762

\bibitem{Coleman:1988}
S. Coleman, ``\textit{Aspects of Symmetry},'' Cambridge University Press, Cambridge, England, 1988. P416.

\bibitem{Affleck:1980ac}
I.~Affleck,
Phys. Rev. Lett. \textbf{46}, 388 (1981)
doi:10.1103/PhysRevLett.46.388


\bibitem{Linde:1980tt}
A.~D.~Linde,
Phys. Lett. B \textbf{100}, 37-40 (1981)
doi:10.1016/0370-2693(81)90281-1

\bibitem{Linde:1981zj}
A.~D.~Linde,
Nucl. Phys. B \textbf{216}, 421 (1983)
[erratum: Nucl. Phys. B \textbf{223}, 544 (1983)]
doi:10.1016/0550-3213(83)90072-X

\bibitem{Friedberg:1976eg}
  R.~Friedberg and T.~D.~Lee,
  Phys.\ Rev.\ D {\bf 15}, 1694 (1977); R.~Friedberg and T.~D.~Lee,
  Phys.\ Rev.\ D {\bf 16}, 1096 (1977); R.~Friedberg and T.~D.~Lee,
  Phys.\ Rev.\ D {\bf 18}, 2623 (1978).

\bibitem{Zhou:2020bzk}
S.~Zhou, S.~Shu and H.~Mao,
Chin. Phys. C \textbf{45}, no.4, 043104 (2021)
doi:10.1088/1674-1137/abdea7
[arXiv:2006.08298 [hep-ph]].

\bibitem{Shu-Ying:2022wqb}
Z.~Shu-Ying, S.~Wan-Ping, M.~Hong,
Acta Phys. Sin. \textbf{71}, no.21, 211101 (2022)
doi:10.7498/aps.71.20220659

\bibitem{Mao:2013qu}
H.~Mao, T.~Wei and J.~Jin,
Phys. Rev. C \textbf{88}, 035201 (2013)
doi:10.1103/PhysRevC.88.035201
[arXiv:1301.6227 [hep-ph]].

\bibitem{Jin:2015goa}
J.~Jin and H.~Mao,
Phys. Rev. C \textbf{93}, no.1, 015202 (2016)
doi:10.1103/PhysRevC.93.015202
[arXiv:1508.03920 [hep-ph]].

\bibitem{Scavenius:2000bb}
O.~Scavenius, A.~Dumitru, E.~S.~Fraga, J.~T.~Lenaghan and A.~D.~Jackson,
Phys. Rev. D \textbf{63}, 116003 (2001)
doi:10.1103/PhysRevD.63.116003
[arXiv:hep-ph/0009171 [hep-ph]].


\bibitem{Palhares:2010be}
L.~F.~Palhares and E.~S.~Fraga,
Phys. Rev. D \textbf{82}, 125018 (2010)
doi:10.1103/PhysRevD.82.125018
[arXiv:1006.2357 [hep-ph]].

\bibitem{Kroff:2014qxa}
D.~Kroff and E.~S.~Fraga,
Phys. Rev. D \textbf{91}, no.2, 025017 (2015)
doi:10.1103/PhysRevD.91.025017
[arXiv:1409.7026 [hep-ph]].

\bibitem{Mintz:2012mz}
B.~W.~Mintz, R.~Stiele, R.~O.~Ramos and J.~Schaffner-Bielich,
Phys. Rev. D \textbf{87}, no.3, 036004 (2013)
doi:10.1103/PhysRevD.87.036004
[arXiv:1212.1184 [hep-ph]].

\bibitem{Stiele:2016cfs}
R.~Stiele and J.~Schaffner-Bielich,
Phys. Rev. D \textbf{93}, no.9, 094014 (2016)
doi:10.1103/PhysRevD.93.094014
[arXiv:1601.05731 [hep-ph]].

\bibitem{Wan-Ping:2019ifh}
Shen Wan-Ping, You Shi-Jia, Mao Hong,
Acta Phys. Sin. \textbf{68}, no.18, 181101 (2019)
doi:10.7498/aps.68.20190798

\bibitem{Skokov:2010sf}
V.~Skokov, B.~Friman, E.~Nakano, K.~Redlich and B.~J.~Schaefer,
Phys. Rev. D \textbf{82}, 034029 (2010)
doi:10.1103/PhysRevD.82.034029
[arXiv:1005.3166 [hep-ph]].

\bibitem{Gupta:2011ez}
U.~S.~Gupta and V.~K.~Tiwari,
Phys. Rev. D \textbf{85}, 014010 (2012)
doi:10.1103/PhysRevD.85.014010
[arXiv:1107.1312 [hep-ph]].

\bibitem{Herbst:2013ail}
T.~K.~Herbst, J.~M.~Pawlowski and B.~J.~Schaefer,
Phys. Rev. D \textbf{88}, no.1, 014007 (2013)
doi:10.1103/PhysRevD.88.014007
[arXiv:1302.1426 [hep-ph]].



\bibitem{Scavenius:2000qd}
O.~Scavenius, A.~Mocsy, I.~N.~Mishustin and D.~H.~Rischke,
Phys. Rev. C \textbf{64}, 045202 (2001)
doi:10.1103/PhysRevC.64.045202
[arXiv:nucl-th/0007030 [nucl-th]].

\bibitem{ParticleDataGroup:2022pth}
R.~L.~Workman \textit{et al.} [Particle Data Group],
PTEP \textbf{2022}, 083C01 (2022)
doi:10.1093/ptep/ptac097

\bibitem{Mustafa:2022got}
M.~G.~Mustafa,
Eur. Phys. J. ST \textbf{232}, no.9, 1369-1457 (2023)
doi:10.1140/epjs/s11734-023-00868-8
[arXiv:2207.00534 [hep-ph]].

\bibitem{Quiros:1999jp}
M.~Quiros,
[arXiv:hep-ph/9901312 [hep-ph]].

\bibitem{Laine:2016hma}
M.~Laine and A.~Vuorinen,
Lect. Notes Phys. \textbf{925}, pp.1-281 (2016)
Springer, 2016,
doi:10.1007/978-3-319-31933-9
[arXiv:1701.01554 [hep-ph]].


\bibitem{Li:2018rfu}
Y.~Li, J.~Hu and H.~Mao,
Phys. Rev. C \textbf{97}, no.5, 054313 (2018)
doi:10.1103/PhysRevC.97.054313
[arXiv:1801.09202 [hep-ph]].

\bibitem{Chen:2015dra}
J.~W.~Chen, J.~Deng, H.~Kohyama and L.~Labun,
Phys. Rev. D \textbf{93}, no.3, 034037 (2016)
doi:10.1103/PhysRevD.93.034037
[arXiv:1509.04968 [hep-ph]].

\bibitem{Fan:2016ovc}
W.~Fan, X.~Luo and H.~S.~Zong,
Int. J. Mod. Phys. A \textbf{32}, no.11, 1750061 (2017)
doi:10.1142/S0217751X17500610
[arXiv:1608.07903 [hep-ph]].

\bibitem{Friman:2011pf}
B.~Friman, F.~Karsch, K.~Redlich and V.~Skokov,
Eur. Phys. J. C \textbf{71}, 1694 (2011)
doi:10.1140/epjc/s10052-011-1694-2
[arXiv:1103.3511 [hep-ph]].

\bibitem{Friman:2014cua}
B.~Friman,
Nucl. Phys. A \textbf{928}, 198-208 (2014)
doi:10.1016/j.nuclphysa.2014.04.012
[arXiv:1404.7471 [nucl-th]].

\bibitem{Gao:2020qsj}
F.~Gao and J.~M.~Pawlowski,
Phys. Rev. D \textbf{102}, no.3, 034027 (2020)
doi:10.1103/PhysRevD.102.034027
[arXiv:2002.07500 [hep-ph]].

\bibitem{Fu:2021oaw}
W.~j.~Fu, X.~Luo, J.~M.~Pawlowski, F.~Rennecke, R.~Wen and S.~Yin,
Phys. Rev. D \textbf{104}, no.9, 094047 (2021)
doi:10.1103/PhysRevD.104.094047
[arXiv:2101.06035 [hep-ph]].

\bibitem{Fu:2023lcm}
W.~j.~Fu, X.~Luo, J.~M.~Pawlowski, F.~Rennecke and S.~Yin,
[arXiv:2308.15508 [hep-ph]].

\bibitem{Fu:2022gou}
W.~j.~Fu,
Commun. Theor. Phys. \textbf{74}, no.9, 097304 (2022)
doi:10.1088/1572-9494/ac86be
[arXiv:2205.00468 [hep-ph]].

\bibitem{Bessa:2008nw}
A.~Bessa, E.~S.~Fraga and B.~W.~Mintz,
Phys. Rev. D \textbf{79}, 034012 (2009)
doi:10.1103/PhysRevD.79.034012
[arXiv:0811.4385 [hep-ph]].

\bibitem{Weinberg:2012pjx}
E.~J.~Weinberg,
Cambridge University Press, 2012,
ISBN 978-0-521-11463-9, 978-1-139-57461-7, 978-0-521-11463-9, 978-1-107-43805-7
doi:10.1017/CBO9781139017787

\bibitem{Masoumi:2016wot}
A.~Masoumi, K.~D.~Olum and B.~Shlaer,
JCAP \textbf{01}, 051 (2017)
doi:10.1088/1475-7516/2017/01/051
[arXiv:1610.06594 [gr-qc]].

\bibitem{Oertel:2008wr}
  M.~Oertel and M.~Urban,
  Phys.\ Rev.\ D {\bf 77}, 074015 (2008)
  [arXiv:0801.2313 [nucl-th]].
  
\bibitem{Garcia:2013eaa}
  A.~F.~Garcia and M.~B.~Pinto,
  Phys.\ Rev.\ C {\bf 88}, no. 2, 025207 (2013)
  [arXiv:1306.3090 [hep-ph]].

\bibitem{Ke:2013wga}
  W.~y.~Ke and Y.~x.~Liu,
  Phys.\ Rev.\ D {\bf 89}, no. 7, 074041 (2014)
  [arXiv:1312.2295 [hep-ph]].
 
\bibitem{Fraga:2018cvr}
  E.~S.~Fraga, M.~Hippert and A.~Schmitt,
  Phys.\ Rev.\ D {\bf 99}, no. 1, 014046 (2019)
  [arXiv:1810.13226 [hep-ph]].
  
\bibitem{Floerchinger:2012xd}
S.~Floerchinger and C.~Wetterich,
Nucl. Phys. A \textbf{890-891}, 11-24 (2012)
doi:10.1016/j.nuclphysa.2012.07.009
[arXiv:1202.1671 [nucl-th]].
 
\bibitem{Drews:2013hha}
M.~Drews, T.~Hell, B.~Klein and W.~Weise,
Phys. Rev. D \textbf{88}, no.9, 096011 (2013)
doi:10.1103/PhysRevD.88.096011
[arXiv:1308.5596 [hep-ph]].

\bibitem{McLerran:2007qj}
L.~McLerran and R.~D.~Pisarski,
Nucl. Phys. A \textbf{796}, 83-100 (2007)
doi:10.1016/j.nuclphysa.2007.08.013
[arXiv:0706.2191 [hep-ph]].

\bibitem{Fukushima:2013rx}
K.~Fukushima and C.~Sasaki,
Prog. Part. Nucl. Phys. \textbf{72}, 99-154 (2013)
doi:10.1016/j.ppnp.2013.05.003
[arXiv:1301.6377 [hep-ph]].

\bibitem{He:2005nk}
L.~y.~He, M.~Jin and P.~f.~Zhuang,
Phys. Rev. D \textbf{71}, 116001 (2005)
doi:10.1103/PhysRevD.71.116001
[arXiv:hep-ph/0503272 [hep-ph]].

\bibitem{Adhikari:2018cea}
P.~Adhikari, J.~O.~Andersen and P.~Kneschke,
Phys. Rev. D \textbf{98}, no.7, 074016 (2018)
doi:10.1103/PhysRevD.98.074016
[arXiv:1805.08599 [hep-ph]].


\bibitem{Rajagopal:2000wf}
K.~Rajagopal and F.~Wilczek,
doi:10.1142/9789812810458\_0043
[arXiv:hep-ph/0011333 [hep-ph]].

\bibitem{Alford:2007xm}
M.~G.~Alford, A.~Schmitt, K.~Rajagopal and T.~Sch\"afer,
Rev. Mod. Phys. \textbf{80}, 1455-1515 (2008)
doi:10.1103/RevModPhys.80.1455
[arXiv:0709.4635 [hep-ph]].

\bibitem{Buballa:2014tba}
M.~Buballa and S.~Carignano,
Prog. Part. Nucl. Phys. \textbf{81}, 39-96 (2015)
doi:10.1016/j.ppnp.2014.11.001
[arXiv:1406.1367 [hep-ph]].

\bibitem{Ma:2019ery}
Y.~L.~Ma and M.~Rho,
Prog. Part. Nucl. Phys. \textbf{113}, 103791 (2020)
doi:10.1016/j.ppnp.2020.103791
[arXiv:1909.05889 [nucl-th]].

\bibitem{Baym:2017whm}
G.~Baym, T.~Hatsuda, T.~Kojo, P.~D.~Powell, Y.~Song and T.~Takatsuka,
Rept. Prog. Phys. \textbf{81}, no.5, 056902 (2018)
doi:10.1088/1361-6633/aaae14
[arXiv:1707.04966 [astro-ph.HE]].

\bibitem{Schaefer:2004en}
B.~J.~Schaefer and J.~Wambach,
Nucl. Phys. A \textbf{757}, 479-492 (2005)
doi:10.1016/j.nuclphysa.2005.04.012
[arXiv:nucl-th/0403039 [nucl-th]].

\bibitem{Chen:2023cjt}
H.~L.~Chen, Z.~B.~Zhu and X.~G.~Huang,
Phys. Rev. D \textbf{108}, no.5, 054006 (2023)
doi:10.1103/PhysRevD.108.054006
[arXiv:2306.08362 [hep-ph]].

\bibitem{Cornwall:1974vz}
J.~M.~Cornwall, R.~Jackiw and E.~Tomboulis,
Phys. Rev. D \textbf{10}, 2428-2445 (1974)
doi:10.1103/PhysRevD.10.2428

\bibitem{Petropoulos:2004bt}
N.~Petropoulos,
[arXiv:hep-ph/0402136 [hep-ph]] and references therein.

\bibitem{Mao:2013gva}
H.~Mao,
Nucl. Phys. A \textbf{925}, 185-198 (2014)
doi:10.1016/j.nuclphysa.2014.02.011
[arXiv:1305.4329 [hep-ph]].

\end{thebibliography}
\end{document}